\documentclass[reprint,
superscriptaddress,
amsmath,amssymb,
aps,
prd]{revtex4-2}

\usepackage{graphicx,wrapfig,lipsum}
\usepackage{subeqnarray}
\usepackage{url,hyperref} 
\usepackage{color}
\usepackage{ulem}
\usepackage{verbatim}
\newcommand{\req}[1]{Eq.\,({\ref{#1}})}
\newcommand{\reqs}[2]{Eqs.\,({\ref{#1}},{\ref{#2}})}

\begin{document}
\title{Covariant Cherenkov Radiation and its Friction Force}
\author{Will Price}
\email{wprice@arizona.edu}
\affiliation{Department of Physics, The University of Arizona, 85721 Tucson, AZ, USA}
\author{Martin S. Formanek}
\email{martin.formanek@eli-beams.eu}
\affiliation{ELI Beamlines Facility, The Extreme Light Infrastructure ERIC, 252 41 Dolní Břežany, Czech Republic}
\author{Johann Rafelski}
\email{johannr@arizona.edu}
\affiliation{Department of Physics, The University of Arizona, 85721 Tucson, AZ, USA}

\date{April 17, 2026}

\begin{abstract}
We derive the covariant generalization of the Frank-Tamm formula describing the Cherenkov radiation by a charged particle moving uniformly with a speed faster than the local speed of light within a homogeneous dielectric medium.  We  use our result to derive the covariant Cherenkov radiation reaction force and obtain a four-force explicitly orthogonal to particle four-velocity consistent with a relativistic friction force. We present the photon emission spectrum that is dependent primarily on the dielectric properties of the medium. We hint at a possible use of this work to interpret an excess of soft photons seen in relativistic hadron collisions. 
\end{abstract}

\maketitle
\section{Introduction}
Unlike radiation emission in vacuum (i.e. in space devoid of matter \cite{fried2012vacuum}), which is associated with particle acceleration~\cite{larmor1897lxiii}, the Cherenkov radiation emission~\cite{Frank:1937fk,Tamm:1991avf} occurs when a charged particle in uniform motion travels through a dielectric medium, at a speed greater than the in-medium speed of light. We offer here a fully frame independent study of this phenomenon. This generalization allows us to obtain a covariant form of the Cherenkov radiation friction (CRF) force due to emission of Cherenkov radiation. We treat the CRF effect at leading order, not considering the resulting deviation of the particle trajectory, from the uniform motion.
 
In prior work Cherenkov radiation was obtained in the rest frame of the medium; we generalize to a frame of reference independent formulation describing the medium motion by a timelike four-velocity $\eta^\mu $. The radiating particle has four-velocity $u^\mu$. The relative Lorentz factor $\Gamma$ and normalized relative speed $\mathcal{V}$ can be formed using the invariant $u \cdot \eta$ and speed of light $c$
\begin{equation}\label{eq:invariant_velocity}
\begin{split}
 \Gamma &\equiv \frac{u\cdot\eta}{c^2} \equiv (1-\mathcal V^2)^{-1/2} \,,\\
 \mathcal V &= \sqrt{1-\Gamma^{-2}} \, .
\end{split}
\end{equation}
In the medium rest frame $\left. \eta^\mu\right|_\text{M} = (c,0,0,0)$ one has $\left.\Gamma\right|_\text{M}=\gamma$ the Lorentz-factor of the particle, and the usual relativistic relations between $\gamma$ and the particle speed $\left. {\mathcal{V}}\right|_\text{M} = {\beta}$ are easily verified: $\gamma = (1-\beta^2)^{-1/2}$. Throughout this work, we use SI units and the Minkowski flat-space metric $g_{\mu\nu} = \text{diag}(1,-1,-1,-1)$.

We consider an isotropic dielectric with no magnetic or magneto-electric properties. The dielectric serves as the scattering  medium  that catalyzes the coherent radiation formation induced by the superluminal charged particle. Coherent refers to many scattering centers that act jointly to form the dielectric medium. The emission of Cherenkov radiation by a single moving particle depends on this coherent optical response of the medium characterized by an index of refraction $n$. In our approach $n$ is a function of generalized invariant energy, converted here to the invariant wavenumber~\cite{Melrose:qu2008}, 
\begin{equation}
\widetilde{k} \equiv \frac{k \cdot \eta}{c} \, ,
\end{equation}
where $k_\mu$ is the wave four-vector of the electromagnetic field. We assume a medium with positive index of refraction $n(\widetilde{k}) > 0$, without absorption $\text{Im}(n) = 0$, and with a regular dispersion relation $dn(\widetilde{k}) / d\widetilde{k} > 0$. 

Our key result presented in the following Section~\ref{sec:summary} is the demonstration that the CRF force has a covariant structure coinciding with the form discussed for covariant material friction~\cite{dunkel2009relativistic, Formanek:2020zwc} that is transverse to the particle 4-velocity $u^\mu$. Therefore, CRF does not suffer from the problems of covariant Larmor radiation friction (LRF) force~\cite{barut1980electrodynamics} for a particle undergoing four-acceleration $a^\mu$ in vacuum. These difficulties in the covariant generalization of the Larmor radiation-friction force arise because the Larmor force acts in a direction parallel to the four-velocity $u^\mu$. Both Larmor and Cherenkov radiation describe classical electromagnetic radiation emitted by a charged particle, with the emitted field carrying away the particle’s energy and momentum. The difference is that Larmor radiation arises from acceleration and is present even in vacuum, whereas Cherenkov radiation arises from superluminal motion relative to a medium so that it depends on the macroscopic medium response through the index of refraction $n$ and the medium four-velocity $\eta^\mu$.

This paper is organized as follows: For the choice of coordinate system and related variable definitions, we refer to Appendix~\ref{sec:coords}. In Section~\ref{sec:summary} the results are summarized and the following sections outline their derivation in detail. In particular:
Section~\ref{sec:constitutive_relations} discusses the covariant form of the constitutive field relations and specifies the form of the constitutive tensor density for isotropic dielectric. Section~\ref{sec:field} evaluates the electromagnetic and displacement field of a uniformly moving charged particle in a dielectric medium. Finally, Section~\ref{sec:RR} derives the covariant form of the radiation friction force that the particle experiences due to Cherenkov radiation and the photon emission spectrum. Appendices \ref{app:inverse}-\ref{app:cylinder_caps} provide definitions and details of mathematical derivations.

\section{Summary of results}\label{sec:summary}

In this work, the Cherenkov radiation friction (CRF) force four-vector acting on a uniformly moving particle in a dielectric medium is derived [for details, see the derivation in Sections \ref{sec:constitutive_relations} - \ref{sec:RR} below, and the final result obtained in~\req{eq:particle_momentum}]
\begin{equation}
\mathcal F^\mu_\mathrm{CRF} = r_\mathrm{CRF}R^{\mu\nu}u_\nu \,,
\end{equation}
where the radiation friction force is orthogonal to the four-velocity of the particle $u^\mu$ 
\begin{equation}\label{eq:Ortho}
\mathcal F^\mu_\mathrm{CRF} u_\mu=0\,,
\end{equation}
since $R^{\mu\nu}$ is an antisymmetric tensor constructed from the two four-velocities $u^\mu$ and $\eta^\mu$
\begin{equation}
 R^{\mu\nu} \equiv \frac{1}{c^2}\left(\eta^\mu u^\nu - u^\mu \eta^\nu\right)\,.
\end{equation}
The orthogonality to the particle four-velocity is a constraint that must be obeyed by all relativistic forces to preserve the on-shell condition $p^2 = m^2u^2=m^2c^2$. It is worth noting that the covariant form of the CRF force coincides with the prescription for the material friction force in a medium~\cite{dunkel2009relativistic, Formanek:2020zwc}. 

Finally, $r_\mathrm{CRF}$ is a coefficient characterizing the strength of the Cherenkov radiation friction
\begin{equation} \label{eq:Ch_coefficient}
\begin{split}
 r_\mathrm{CRF} = \frac{q^2\mu_0 c}{4\pi \mathcal V\Gamma}&\int_0^\infty d\widetilde{k} \, \widetilde{k}\left(1-\frac{1}{n^2(\widetilde{k})\mathcal V^2}\right)\\
 &\times \Theta[n(\widetilde{k})\mathcal V-1] \,,
\end{split}
\end{equation}
where $q$ is the particle charge and $\mu_0$ is the vacuum permeability. The step function in~\req{eq:Ch_coefficient} ensures that Cherenkov radiation is only emitted when the relative velocity between particle and medium is faster than the local speed of light, i.e. $n(\widetilde{k})\mathcal V>1$.

Our covariant result for the CRF force is in agreement with the Frank-Tamm formula for the stopping power of the particle evaluated in the medium rest frame~\cite{Tamm:1991avf} (see Section~\ref{sec:RR}).
\begin{equation}\label{eq:Frank-Tamm}
\begin{split}
 \left. \frac{dE}{dx}\right|_\text{M} = -\frac{q^2\mu_0}{4\pi} &\int_0^\infty d\omega\, \omega \left(1 - \frac{1}{n^2(\omega)\beta^2}\right)\\
 &\times \Theta[n(\omega)\beta -1]\,,
\end{split}
\end{equation}
where $\left. dE/dx\right|_\text{M}$ is the particle energy loss per unit distance for an observer in the rest frame of the medium and $\omega$ is the angular frequency of the radiation. 

While the CRF force preserves the on-shell condition $p^2 = m^2u^2=m^2c^2$, this is not true for the Larmor radiation friction (LRF) force~\cite{larmor1897lxiii} which has a covariant generalization for a particle that undergoes four-acceleration $a^\mu$ in vacuum~\cite{barut1980electrodynamics}
 \begin{equation}
 \mathcal F^\mu_\mathrm{LRF} = \frac{2}{3}\frac{q^2\mu_0}{4\pi c^3} a^2 u^\mu \, .
 \end{equation}
 The LRF force is manifestly parallel to the particle four-velocity. The Larmor force must then be supplemented with an ad-hoc Schott term \cite{Schott:1912em} to satisfy the orthogonality, which was accomplished by the Lorentz-Abraham-Dirac (LAD) equation of motion for the radiation friction force~\cite{lorentz1952}. This formulation comes with well-documented issues concerning initial condition, causality, and runaway solutions~\cite{poisson1999introduction,spohn2004dynamics,Rohrlich:2007cl}. Clearly, the classical LRF force is not complete, and there have been many proposals to amend it \cite{caldirola1956new,Landau:1975pou,Ford:ra1993,Gralla:2009md,Price:2021zqq}. See also recent classical and quantum reviews of this phenomenon \cite{DiPiazza:2011tq,Burton:2014wsa,Blackburn:2019rfv,Gonoskov:2021hwf,Fedotov:2022ely}.

 Strikingly, Cherenkov radiation emitted by a uniformly moving particle creates a force that is naturally orthogonal to the particle four-velocity and thus intrinsically preserves the on-shell condition. It is the presence of the medium four-velocity $\eta^\mu$ that allows us to obtain a consistent CRF force. Such a four-vector is not available in the case of Larmor radiation in the true empty space devoid of matter, the vacuum. However, in our Universe there is no such state: The Universe has a preferred cosmological frame of reference defined by the CMB (cosmic microwave background) while being filled with matter and radiation moving with reference to this frame \cite{Durrer:2015lza,ellis2012relativistic}. Once the radiation-friction force is determined for particles accelerated with respect to this preferred reference frame of the Universe, defined by the four-vector $\eta^\mu$ of the CMB background, we could infer the corresponding universal radiation-friction force. In general, this force may differ from the vacuum Larmor result, precisely because the additional four-vector $\eta^\mu$ enters the dynamics.

We can estimate the strength of the Cherenkov radiation friction force for an ultrarelativistic electron (charge $e<0$, mass $m$, $\mathcal{V} \approx 1$) in water with index of refraction $n \approx 1.33$, approximately constant over the water transparency band. Then the magnitude of invariant Cherenkov acceleration $|a_\text{CRF}| \equiv \sqrt{-a_\text{CRF}^2}$ which corresponds to the magnitude of the three-acceleration experienced by the electron in its co-moving frame $\left.a^\mu_\text{CRF}\right|_\text{C} = (0,\boldsymbol{a}_c)$, $|a_\text{CRF}| = |\boldsymbol{a}_c| $ is
\begin{equation}\label{aCRF1}
|a_\text{CRF}| = \frac{|\mathcal{F}_\text{CRF}|}{m} = \frac{\sqrt{-\mathcal{F}_\text{CRF}^2}}{m} = \frac{e^2 \mu_0}{4\pi m}J\\,
\end{equation}
where the integral $J$ is approximately
\begin{equation}\label{aCRF2}
\begin{split}
J &\approx \int_0^\infty d\omega\, \omega \left(1 - \frac{1}{n^2}\right)\\
&\approx \frac{1}{2}\left(1 - \frac{1}{n^2}\right) \left(\omega_\text{max}^2-\omega_\text{min}^2\right)\approx 2 \cdot 10^{31}\, \text{s}^{-2}
\end{split}
\end{equation}
where the integration over all frequencies was replaced by frequencies in the typical range of water transparency 200 - 800 nm ($\omega_\text{min} = 2.4\cdot 10^{15}$ Hz, $\omega_\text{max} = 9.4 \cdot 10^{15}$ Hz) \cite{hale1973optical}. 

Outside the transparency region of water, absorption is significant and the present model no longer applies, since we assume a real index of refraction. The electromagnetic waves emitted in spectral region outside of optical transparency window are attenuated, so their contribution to the propagating Cherenkov flux is suppressed. Emission at these frequencies may nevertheless contribute to the particle’s energy loss through dissipation in the medium, e.g. as heating. The treatment of absorptive media lies beyond the scope of the present work but possible extension in this direction is briefly discussed in Appendices \ref{app:contour_int} and \ref{app:proper_time_int}. 

The classical treatment is justified in this spectral range as the energy of emitted photons is negligible compared to the energy of relativistic electrons and the emitted wavelengths are sufficiently large (by a factor of several hundred) for the medium to be described by a macroscopic refractive index rather than by its microscopic atomic structure. 

The numerical value of the result in \req{aCRF2} is dominated by the value of the UV cutoff. The magnitude of the invariant Cherenkov acceleration evaluated in the water transparency region is then $|a_\text{CRF}| \approx 6 \times 10^{16}$ m/s$^2$. 
 
The full covariant in medium radiation friction force for a superluminal and accelerated charged particle is still left to be formulated \cite{Lynch:2019xfl} (see also discussion in Section \ref{sec:discussion}). For now, the Larmor and Cherenkov radiation forces can be naively compared by evaluating the ratio of their invariant magnitudes
\begin{equation}
\frac{|\mathcal{F}_\text{CRF}|}{|\mathcal{F}_\text{LRF}|} = \frac{3}{2}\frac{c^2 J}{a^2}\,.
\end{equation}
The magnitude of invariant acceleration (originating from any source, including heating of the medium) that the electron in water has to experience in order for both forces to be comparable is
\begin{equation}
|a| \approx \sqrt{\frac{3}{2}c^2 J} \approx 2 \cdot 10^{24}\, \text{m/s}^2 \approx 10^{-5} a_\text{crit}\,. 
\end{equation}
The critical acceleration $a_\text{crit} = mc^3/\hbar = 2.33 \cdot 10^{29}$ m/s$^2$ corresponds to an electron accelerated by the Schwinger critical electric field $E_S=m^2c^3/|e|\hbar$.

\section{Covariant form of the constitutive equations}\label{sec:constitutive_relations}
In a linear medium, the non-covariant constitutive relations between the electromagnetic fields $\mathbf E$ and $\mathbf B$ and excitation fields $\mathbf D$ and $\mathbf H$ are
\begin{align}
 D^i &= \varepsilon^{ij}E^j\,, \\
 B^i &= \mu^{ij}H^j \,,
\end{align}
where the dielectric permittivity $\varepsilon^{ij}$ and the magnetic permeability $\mu^{ij}$ are three-dimensional Cartesian tensors. We assume that $\mu^{ij}$ is invertible so that $\mathbf H$ can be written in terms of $\mathbf B$ as
\begin{equation}
 H^i = (\mu^{-1})^{ij}B^j \, .
\end{equation}
Additionally, a material medium could possess a magneto-electric coupling, and the most general form of linear constitutive equations is~\cite{ODell:1962th}
\begin{align}
 \label{eq:constitutive_eqn1} D^i &= \varepsilon^{ij} E^j + \zeta^{ij} B^j\,, \\
 \label{eq:constitutive_eqn2} H^i &= (\mu^{-1})^{ij} B^j + (\zeta^\dagger)^{ij} E^j \, ,
\end{align}
where $\zeta^{ij}$ is the magneto-electric coupling tensor.

Let us define the covariant electromagnetic tensor $F^{\mu\nu}$ and the displacement tensor $\mathcal H^{\mu\nu}$
\begin{align}
 F^{\mu\nu} &\equiv \left(\begin{array}{cc}
 0 & -\pmb{E}/c \\
 \pmb{E}/c & -\epsilon_{ijk}B^k 
 \end{array} \right),\\ 
 \mathcal{H}^{\mu\nu} &\equiv \left(\begin{array}{cc}
 0 & -\pmb{D}c \\
 \pmb{D}c & -\epsilon_{ijk}H^k 
 \end{array}
 \right),
\end{align}
where $\epsilon_{123} = +1$ is the three dimensional Levi-Civita tensor. We can formulate a covariant version of the constitutive relations by introducing a rank-four constitutive tensor density $\hat{z}_{\mu\nu\alpha\beta}$ such that~\cite{Post:1997fo,price2025covariant}
\begin{equation}\label{eq:covariant_constitutive_eqn}
 \mathcal{H}_{\mu\nu}(x) = \frac{1}{2}\int d^4x' \, \hat{z}_{\mu\nu\alpha\beta}(x-x')F^{\alpha\beta}(x') \,.
\end{equation}
This formulation allows for non-local dependence of $H^{\mu\nu}$ on $F^{\mu\nu}$. For local behavior, the density has to be in the form
\begin{equation}
 \hat{z}_{\mu\nu\alpha\beta}(x - x') = Z_{\mu\nu\alpha\beta}(x') \delta^4(x - x')\,,
\end{equation}
where $Z_{\mu\nu\alpha\beta}(x)$ is the local constitutive tensor and for this local case the constitutive relations read
\begin{equation} 
 \mathcal{H}_{\mu\nu}(x) = \frac{1}{2}Z_{\mu\nu\alpha\beta}(x)F^{\alpha\beta}(x)\,.
\end{equation}
Finally, in Maxwell theory in vacuum, the covariant constitutive equations simply reduce to
\begin{equation}
 \mathcal H^{\mu\nu} = \frac{1}{\mu_0}F^{\mu\nu} \,.
\end{equation}

By comparing the components of the covariant and non-covariant form of the constitutive relations, the components of $\hat{z}_{\mu\nu\alpha\beta}$ can be related to $\hat{\varepsilon}^{ij}$, $\hat{\mu}^{ij}$, and $\hat{\zeta}^{ij}$
\begin{align}
 \label{eq:epsilon_noncovariant}c^2\hat{\varepsilon}^{ij} &= -\hat{z}^{i0j0} \,,\\
 \label{eq:mu_noncovariant} (\hat{\mu}^{-1})^{ij} &= -\frac{1}{4}\epsilon^{ikl}\hat{z}_{klmn}\epsilon^{mnj}\,,\\
 \hat{\zeta}^{ij} &= -\frac{1}{2}\hat{z}^{i0}_{\,\,\,\,\,\,lk}\epsilon^{jlk} \,.
\end{align}
We can now generalize the three-dimensional quantities to four-dimensional forms \cite{Schuster:2017mdx} by introducing an observer with four-velocity $\eta^\mu$ satisfying $\eta^2=c^2$
\begin{align}
 \label{eq:epsilon} c^2\hat{\varepsilon}^{\mu\nu} &= -\hat{z}^{\mu\alpha\nu\beta}\eta_\alpha \eta_\beta/c^2\,,\\
 \label{eq:mu} (\hat{\mu}^{-1})^{\mu\nu} &= \frac{1}{4}\epsilon^{\mu\rho\alpha\beta}\hat{z}_{\alpha\beta\gamma\delta}\epsilon^{\nu\sigma\gamma\delta}\eta_\rho \eta_\sigma/c^2\,,\\
 \hat{\zeta}^{\mu\nu} &= \frac{1}{2}\hat{z}^{\mu\rho\alpha\beta}\epsilon_{\alpha\beta}^{\quad\nu\sigma}\eta_\rho \eta_\sigma /c^2\,.
\end{align}
where $\epsilon_{0123} = +1$ is the four-dimensional Levi-Civita tensor. Due to the symmetry properties of the displacement and electromagnetic field tensors, the constitutive tensor density obeys the following symmetries~\cite{Post:1997fo}
\begin{equation}
 \hat{z}_{\mu\nu\alpha\beta} = -\hat{z}_{\nu\mu\alpha\beta} = -\hat{z}_{\mu\nu\beta\alpha} = \hat{z}_{\alpha\beta\mu\nu} \,.
\end{equation}
As a consequence and due to antisymmetric properties of the Levi-Civita tensor, each tensor density is orthogonal to the observer's velocity
\begin{equation}
 \hat{\varepsilon}^{\mu\nu}\eta_\nu = (\hat{\mu}^{-1})^{\mu\nu}\eta_\nu = \hat{\zeta}^{\mu\nu}\eta_\nu = 0 \,,
\end{equation}
and the permittivity and permeability tensor densities are both symmetric
\begin{equation}
 \hat{\varepsilon}^{\mu\nu} = \hat{\varepsilon}^{\nu\mu} \quad , \quad (\hat{\mu}^{-1})^{\mu\nu} = (\hat{\mu}^{-1})^{\nu\mu} \, .
\end{equation}
The magneto-electric tensor density $\hat{\zeta}^{\mu\nu}$ is neither symmetric nor antisymmetric. 

Conversely, it is also possible to express the constitutive tensor density $\hat{z}^{\mu\nu\alpha\beta}$ in terms of the three tensor densities $\hat{\varepsilon}^{\mu\nu}$, $(\hat{\mu}^{-1})^{\mu\nu}$, and $\hat{\zeta}^{\mu\nu}$~\cite{bel2000radiation}
\begin{equation} \label{eq:bel}
\begin{split}
 c^2 \hat{z}^{\mu\nu\alpha\beta} &= c^2\hat{\varepsilon}^{\mu\beta}\eta^\nu \eta^\alpha + c^2\hat{\varepsilon}^{\nu\alpha}\eta^\mu \eta^\beta - c^2\hat{\varepsilon}^{\mu\alpha}\eta^\nu \eta^\beta \\ 
 &- c^2\hat{\varepsilon}^{\nu\beta}\eta^\mu \eta^\alpha + \epsilon^{\mu\nu}_{\quad\rho\sigma}\epsilon^{\alpha\beta}_{\quad\gamma\delta}(\hat{\mu}^{-1})^{\sigma\delta}\eta^\rho \eta^\gamma \\ 
 &+ \epsilon^{\mu\nu}_{\quad\gamma\delta}\left[(\hat{\zeta}^\dagger)^{\delta\alpha}\eta^\beta \eta^\gamma - (\hat{\zeta}^\dagger)^{\delta\beta}\eta^\alpha \eta^\gamma\right] \\ 
 &+ \epsilon^{\alpha\beta}_{\quad\gamma\delta}\left(\hat{\zeta}^{\mu\delta}\eta^\nu \eta^\gamma - \hat{\zeta}^{\nu\delta}\eta^\mu \eta^\gamma\right) \, .
 \end{split}
\end{equation}

A linear, isotropic, homogeneous medium can be characterized by an index of refraction, which is given by two dielectric parameters; medium permittivity and permeability. We can form analogous Lorentz invariant versions of these constants by taking the trace of the dielectric tensors. For example, the scalar permittivity density $\hat{\varepsilon}$ can be defined as 
\begin{equation}
 \hat{\varepsilon}(x-x') \equiv -\frac{1}{3}g_{\mu\nu}\hat{\varepsilon}^{\mu\nu}(x-x') \, .
\end{equation}
For an isotropic medium, the permittivity tensor density can be written as
\begin{equation}
 \hat{\varepsilon}^{\mu\nu}(x-x') = -\hat{\varepsilon}(x-x') P^{\mu\nu} \,,
\end{equation}
where the projector $P^{\mu\nu}$ on the subspace orthogonal to $\eta^\mu$ is defined in \req{eq:projector}. This form of the permittivity tensor density is diagonal for an observer in the medium rest frame with a vanishing $\hat{\varepsilon}^{00}$ component, as should be the case for an isotropic medium. Using this representation, the constitutive tensor density can be written as 
\begin{equation} \label{eq:constitutive_tensor}
\begin{split}
 c^2 \hat{z}^{\mu\nu\alpha\beta} &= -\hat{\varepsilon} c^2(P^{\mu\beta}\eta^\nu \eta^\alpha + P^{\nu\alpha} \eta^\mu \eta^\beta \\ 
 &- P^{\mu\alpha}\eta^\nu\eta^\beta - P^{\nu\beta}\eta^\mu\eta^\alpha) \\ 
 &- \hat{\mu}_0^{-1} g_{\delta\sigma}\epsilon^{\mu\nu\rho\sigma}\epsilon^{\alpha\beta\gamma\delta}\eta_\rho \eta_\gamma \, ,
\end{split}
\end{equation}
where we have assumed that the medium has no magnetic or magnetoelectric properties, so that the permeability density is given by
\begin{align}
 (\hat{\mu}^{-1})^{\mu\nu}(x-x') &= - \hat{\mu}_0^{-1}(x-x') P^{\mu\nu}\,,\label{eq:perm1}\\
 \hat{\mu}_0^{-1}(x-x') &= \frac{1}{\mu_0}\delta^4(x-x')\label{eq:perm2}\,,
\end{align}
and the magneto-electric coupling tensor density is zero
\begin{equation}
 \hat{\zeta}^{\mu\nu} = 0\,.
\end{equation}

\section{Field of a uniformly moving particle in an isotropic medium}\label{sec:field}
The field equations for a particle passing through the medium can now be solved. Maxwell's equations within matter are~\cite{Jackson:1998nia} 
\begin{align}
 \label{eq:field_eqn1} \partial_\mu \mathcal H^{\mu\nu} = j^\nu \\ 
 \label{eq:field_eqn2} \partial_\mu \widetilde F^{\mu\nu} = 0 \,,
\end{align}
where $\widetilde{F}_{\mu\nu} \equiv \epsilon_{\mu\nu\alpha\beta}F^{\alpha\beta}/2$ is the dual electromagnetic tensor. If the index of refraction of the medium is dependent on the wavenumber (frequency), the constitutive relation must be nonlocal and take the form of a convolution integral \req{eq:covariant_constitutive_eqn}. Then, by the convolution theorem, the linear constitutive relations hold in the Fourier space
\begin{equation}
 \mathcal H_{\mu\nu}(k) = \frac{1}{2}\hat{z}_{\mu\nu\alpha\beta}(k)F^{\alpha\beta}(k) \, ,
\end{equation}
where the four-dimensional inverse Fourier transform of a function $f(k)$ is defined as 
\begin{equation}
 f(x) \equiv \int \frac{d^4k}{(2\pi)^4} \,f(k) e^{-ik\cdot x} \, .
\end{equation}
Since in the Fourier space the electromagnetic tensor is
\begin{equation}
 F^{\mu\nu}(k) = -i [k^\mu A^\nu(k) - k^\nu A^\mu(k)]\,,
\end{equation}
the two field equations~\req{eq:field_eqn1} and~\req{eq:field_eqn2} can be expressed instead as a single equation for the potential $A^\mu$ 
\begin{equation} \label{eq:field_eqn}
 -\hat{z}^{\mu\nu\alpha\beta}(k)k_\mu k_\alpha A_\beta(k) = j^\nu(k) \, . 
\end{equation}
This field equation is identical to the field equation found by applying
covariant linear response theory to our problem~\cite{Formanek:2021blc}. Let us take the form of the constitutive tensor for an isotropic medium~\req{eq:constitutive_tensor} but treating $\hat{\varepsilon}(k)$ and $\hat{\mu_0}^{-1}(k)$ as Fourier transformed densities. Since the medium considered here does not have magnetic properties \reqs{eq:perm1}{eq:perm2}, the Fourier transformation of the permeability density is
\begin{equation}
 \hat{\mu}_0^{-1}(k) = \mu_0^{-1}
\end{equation}
the usual vacuum permeability. Using \req{eq:constitutive_tensor}, the field equation can be expressed as 
\begin{equation}\label{eq:FT4potential}
\begin{split}
 &\left\{\left(c^2\mu_0^{-1}k^2+[\hat{\varepsilon}(k) c^2-\mu_0^{-1}]\widetilde{k}^2c^2\right) \delta^\mu_\nu - c^2\mu_0^{-1}k^\mu k_\nu + \right.\\ 
 &[\hat{\varepsilon}(k) c^2-\mu_0^{-1}]\left.\left[k^2\eta^\mu \eta_\nu - \widetilde{k}c(k^\mu\eta_\nu + \eta^\mu k_\nu)\right]\right\}A^\nu(k) \\
 &= -c^2j^\mu(k) \, .
\end{split}
\end{equation}
Now we can identify the wavenumber-dependent index of refraction
\begin{equation}
 n(\widetilde{k}) = c \sqrt{\hat{\varepsilon}(\widetilde{k}) \hat{\mu}_0(\widetilde{k})} = c \sqrt{\hat{\varepsilon}(\widetilde{k})\mu_0}\,.
\end{equation}
In this work, we will assume that the index of refraction is real $\text{Im}[n(\widetilde{k})] = 0$, positive $n(\widetilde{k})>0$, and with ordinary dispersion $dn(\widetilde{k})/d\widetilde{k} > 0$. Using the index of refraction simplifies the algebraic expression in \req{eq:FT4potential} for the four-potential $A^\nu(k)$ in the momentum space to
\begin{equation}\label{eq:Av}
\begin{split}
 &\left\{\left[k^2+\left(n^2-1\right)\widetilde{k}^2\right] \delta^\mu_\nu - k^\mu k_\nu + \left(n^2-1\right)\right.\\ 
 &\times\left.\frac{1}{c^2}\left[k^2\eta^\mu \eta_\nu - \widetilde{k}c(k^\mu\eta_\nu + \eta^\mu k_\nu)\right]\right\}A^\nu(k) = - \mu_0 j^\mu(k) \, .
\end{split}
\end{equation}
The determinant of the tensor on the left hand side is equal to zero, which means that the inverse can be performed only on a subspace. Such subspace of $A^\nu$ is spanned by the 4-potentials satisfying the Lorenz gauge $k \cdot A = 0$ where this expression can be inverted (for details see Appendix~\ref{app:inverse}) to obtain the propagator $D^{\mu\nu}$
\begin{equation} \label{eq:potential}
 A^\mu(k) = \mu_0 D^{\mu\nu}(k)j_\nu(k) \, ,
\end{equation}
where the propagator is
\begin{equation} \label{eq:momentum_propagator}
\begin{split}
 D^{\mu\nu}(k) &= \frac{-g^{\mu\nu} + (1-1/n^2)\eta^\mu \eta^\nu /c^2}{k^2 + (n^2-1)\widetilde{k}^2}\\
 &- \frac{(1-1/n^2)\widetilde{k}k^\mu\eta^\nu/c }{k^2 [k^2 + (n^2-1)\widetilde{k}^2]}\, .
\end{split}
\end{equation}

Although the form of the propagator~\req{eq:momentum_propagator} is valid in a specific gauge, this gauge choice is not relevant to our final result, which is gauge invariant. The second term in \req{eq:momentum_propagator} also does not contribute to the electromagnetic field tensor because it constitutes a gauge transformation of the four-potential and will be dropped, see Appendix~\ref{app:inverse}. The first term coincides with the form of photon propagator in material medium formulated by Schwinger et al. \cite{Schwinger:1974rq}.

The position space form of the propagator is given by the inverse Fourier transformation
\begin{equation}
\begin{split}
 D^{\mu\nu}(x-x') = -\int \frac{d^4k}{(2\pi)^4} \, \frac{P_n^{\mu\nu}(k) e^{-ik\cdot(x-x')}}{k^2 + [n^2(k)-1]\widetilde{k}^2} \,,
\end{split}
\end{equation}
where the tensor structure is given by
\begin{equation}\label{eq:Pmn}
 P^{\mu\nu}_n(k) \equiv g^{\mu\nu} - \frac{n^2(k)-1}{n^2(k)}\frac{\eta^\mu \eta^\nu}{c^2}.
\end{equation}
We assume that the index of refraction is only a function of the invariant wavenumber, i.e. $n(k) = n(\widetilde{k})$. It is then possible to use the residue theorem to integrate over the other three momentum integrals (see Appendix~\ref{app:contour_int}). The result for the position-space form of the propagator as an integral over the invariant wavenumber is
\begin{equation} \label{eq:propagator}
\begin{split}
 D^{\mu\nu}(x-x') &= \frac{1}{8\pi^2}\int_{-\infty}^\infty d\widetilde{k} \, P^{\mu\nu}_n(\widetilde k) \\
 &\times \frac{e^{-i\widetilde{k}\left[\widetilde{x}-\widetilde{x}' - n(\widetilde k)\sqrt{-(\widetilde{x}_\perp-\widetilde{x}'_\perp)^2}\right]}}{\sqrt{-(\widetilde{x}_\perp - \widetilde{x}'_\perp)^2}} \,,
\end{split}
\end{equation}
where $\widetilde{x}-\widetilde{x}' = (x-x')\cdot \eta /c$ is the difference in the coordinate along the medium four-velocity and $\widetilde{x}^\mu_\perp - \widetilde{x}'^\mu_\perp = P^{\mu\nu}(x-x')_\nu$ is the difference in the transverse coordinate according to the decomposition in \req{eq:etadecompose}.

With this propagator it is possible to calculate the field of a point particle with current density \cite{Rafelski:2017hyt}
\begin{equation}
 j^\mu(x) = qc \int_{-\infty}^\infty d\tau \, u^\mu(\tau) \delta^4[x-z(\tau)] \,,
\end{equation}
which describes a point particle with a trajectory $z^\mu(\tau)$ and four-velocity $u^\mu(\tau)$ parameterized by its proper time $\tau$. Using the Fourier transformation of ~\req{eq:potential} in the form of a convolution
\begin{equation}
 A^\mu(x) = \mu_0 \int d^4x' D^{\mu\nu}(x-x') j_\nu(x') 
\end{equation}
to calculate the potential yields
\begin{equation} \label{eq:potential2}
\begin{split}
 A^\mu(x) &= \frac{\mu_0 q c}{8\pi^2}\int_{-\infty}^\infty d\widetilde{k}\, P_n^{\mu\nu}(\widetilde k) \int_{-\infty}^\infty d\tau \, u_\nu(\tau) \\ 
 &\times\frac{e^{-i\widetilde{k}\left[\widetilde{x}-\widetilde{z}(\tau) - n(\widetilde k)\sqrt{-[\widetilde{x}_\perp-\widetilde{z}_\perp(\tau)]^2}\right]}}{\sqrt{-[\widetilde{x}_\perp-\widetilde{z}_\perp(\tau)]^2}} \,.
\end{split}
\end{equation}
This expression describes the position-space four-potential for an arbitrary motion of a charged particle through an isotropic, homogeneous medium with an invariant index of refraction $n(\widetilde{k})$. 

Here, we study the simplest case of a uniformly moving particle. Such a particle has in the zeroth order no acceleration and therefore cannot emit Larmor radiation~\cite{Schwinger:1974rq}. The four-velocity $u^\mu$ of the uniformly moving particle is a constant, and therefore its trajectory $z^\mu(\tau)$ is given by
\begin{equation}
 z^\mu(\tau) = u^\mu \tau \,,
\end{equation}
with a choice of initial condition $z^\mu(0) = 0$.

In this simplified case, the proper time integral in~\req{eq:potential2} can be computed (see Appendix~\ref{app:proper_time_int})
\begin{equation}\label{eq:potential_uniform}
\begin{split}
 A^\mu(x) = \frac{i\mu_0 q}{8\pi\mathcal V\Gamma}&\int_{-\infty}^\infty d\widetilde{k} \, P_n^{\mu\nu}(\widetilde k)u_\nu\\
 &\times e^{-i\widetilde{k}(\widetilde{x}-x_\parallel/\mathcal{V})} I(\widetilde{k},x_\perp) \, .
\end{split}
\end{equation}
In this expression, the coordinates $x_\parallel$ and $x_\perp$ defined in \reqs{eq:x_parallel}{eq:x_perp}, were used. Respectively, they correspond to coordinates parallel and perpendicular to the particle motion on the subspace orthogonal to medium four-velocity. Finally, the function $I(\widetilde{k},x_\perp)$ is given by (see Appendix~\ref{app:proper_time_int})
\begin{equation}\label{eq:Iintegralsol}
\begin{split}
 I(\widetilde{k},x_\perp) &\equiv \Theta(1-n\mathcal V)H_0^{(1)}(\xi) + \Theta(n\mathcal V-1) \\ 
 &\times\left[\Theta(\widetilde{k}) H_0^{(1)}(\xi) - \Theta(-\widetilde{k})H_0^{(2)}(\xi)\right] \, ,
\end{split}
\end{equation}
where $H_0^{(1)}$ and $H_0^{(2)}$ are the zeroth order Hankel functions of the first and second kinds, while their argument $\xi$ is defined as
\begin{equation}
 \xi \equiv \frac{|\widetilde{k}|x_\perp}{\mathcal V}\sqrt{n^2\mathcal V^2-1} \, .
\end{equation}
The argument $\xi$ is real for the superluminal particle motion ($n\mathcal V>1$) and imaginary for the subluminal particle motion ($n\mathcal V<1)$. This transition between the real and imaginary arguments around the Cherenkov condition is what mathematically leads to the presence of a radiation field. 

Using the four-potential~\req{eq:potential_uniform} of the field emitted by the uniformly moving particle, we can calculate the corresponding electromagnetic field tensor as $F^{\mu\nu} = \partial^\mu A^\nu - \partial^\nu A^\mu$. The resulting expression is (see Appendix~\ref{app:EMtensor} for derivation)
\begin{equation}\label{eq:emtensor}
\begin{split}
 &F^{\mu\nu}(x) = \frac{i\mu_0q}{4\pi\mathcal V\Gamma}\int_{-\infty}^\infty d\widetilde{k} \, e^{-i\widetilde{k} (\widetilde{x} - x_\parallel /\mathcal V)}\\
 &\times \left[-i\widetilde{k} I \left(1-\frac{1}{n^2\mathcal V^2}\right)\frac{\eta^{[\mu} u^{\nu]}}{c}
 + \partial^{[\mu} (I) v^{\nu]}\right] \,,
\end{split}
\end{equation}
where the four-vector $v^\mu$ is defined as
\begin{equation}
 v^\mu(\widetilde k) \equiv P^{\mu\nu}_n(\widetilde k) u_\nu
\end{equation}
and the square brackets denote an antisymmetric combination of indices
\begin{equation}\label{eq:antisym}
 a^{[\mu}b^{\nu]} \equiv \frac{1}{2}(a^\mu b^\nu - a^\nu b^\mu)\,.
\end{equation}
The displacement tensor can be evaluated using the covariant constitutive relations \req{eq:covariant_constitutive_eqn} as is demonstrated in Appendix~\ref{app:displacementtensor}
\begin{equation}\label{eq:displacementtensor}
\begin{split}
 &\mathcal H^{\mu\nu}(x) = \frac{iq}{4\pi\mathcal V\Gamma}\int d\widetilde{k} \, e^{-i\widetilde{k}(\widetilde{x} - x_\parallel /\mathcal V)}\\
 &\times\left[-i\widetilde{k} I n^2\left(1-\frac{1}{n^2\mathcal V^2}\right)\frac{\eta^{[\mu} u^{\nu]}}{c} + \partial^{[\mu}(I) u^{\nu]}\right] \, .
\end{split}
\end{equation}
Both $F^{\mu\nu}(x)$ and $\mathcal H^{\mu\nu}(x)$ are presented as functions of coordinates $\widetilde{x}$, $x_\parallel$, and the dependence on $x_\perp$ is contained in $I(\widetilde{k},x_\perp)$ \req{eq:Iintegralsol}. 

\section{Covariant form of the Cherenkov radiation friction force}\label{sec:RR}
The infinitesimal momentum change of the radiation field is calculated by integrating the energy-momentum tensor over a spacelike hypersurface surrounding the charge~\cite{Schild:1960on},
\begin{equation}
 dp^\mu_\text{EM} = \frac{1}{c}\int T^{\mu\nu}d^3 \sigma_\nu \, ,
\end{equation}
where $d^3 \sigma_\nu$ represents the hypersurface element. The change in the momentum of the field is generally independent of the choice of the hypersurface. We can then choose the hypersurface to be a curved surface of a hypercylinder centered around the particle \cite{barut1980electrodynamics}
\begin{equation}
 d^3\sigma_\nu = - g_{\nu\alpha}e_\perp^\alpha x_\perp dx_\parallel d\phi d \widetilde{x} \,.
\end{equation}
where $\phi \in (0, 2\pi)$ is the polar angle in the 2-plane orthogonal to both $\eta^\mu$ and $u^\mu$ and $\widetilde{x}$ is the coordinate along the medium four-velocity [\req{eq:etadecompose}]. $x_\parallel$ is the coordinate along the particle four-velocity [\req{eq:x_parallel}] integrated from hypercylinder caps at $-L$ to $L$, for the steady state solution in the limit $L \rightarrow \infty$. For the observer in the medium rest frame $\left. d\widetilde{x}\right|_\text{M} = cdt$ and $-\left.e_\perp^\mu\right|_\text{M}$ is a normal vector that points outward with respect to the curved cylinder surface [\req{eq:e_perp}]. For the steady state situation explored here the contributions from the hypercylinder caps cancel each other, for formal proof see Appendix~\ref{app:cylinder_caps}. The covariant change in field momentum is then given by
\begin{equation}\label{eq:radiated_momentum}
 dp^\mu_\text{EM} = -\frac{x_\perp}{c} \int d\phi dx_\parallel d\widetilde{x}\, T^{\mu\nu}e_{\perp\nu} \,.
\end{equation}

The energy momentum tensor of the electromagnetic field in the medium is given as
\begin{equation}
 T^{\mu\nu} = F^{\mu\alpha}\mathcal{H}_\alpha^{*\nu} + \frac{1}{4}F^{\alpha \beta}\mathcal{H}^*_{\alpha\beta} g^{\mu\nu}\,, 
\end{equation}
where the complex conjugate of the displacement field tensor is taken to be consistent with the usual three-dimensional formulas for the energy and momentum radiated by fields with harmonic time dependence~\cite{Jackson:1998nia}. Let us define an integration kernel $\mathcal{T}^{\mu\nu}(\widetilde{k}, \widetilde{k}', x_\perp)$ containing the tensor structure of the energy momentum tensor
\begin{equation}\label{eq:kernel}
\begin{split}
 T^{\mu\nu}(x) &= \int d\widetilde{k} d\widetilde{k}' \, e^{-i(\widetilde{k}-\widetilde{k}')(\widetilde{x} - x_\parallel/\mathcal{V})}\\
 &\times \mathcal{T}^{\mu\nu}(\widetilde{k},\widetilde{k}',x_\perp) \,.
\end{split}
\end{equation}
Using this kernel, the change of momentum is then given by
\begin{equation}
\begin{split}
 dp^\mu_\text{EM} &= -\frac{x_\perp}{c}\int d\phi dx_\parallel d\widetilde{k} d\widetilde{k}' d\widetilde{x} \, e^{-i(\widetilde{k}-\widetilde{k}')(\widetilde{x}-x_\parallel/\mathcal V)}\\
 &\times \mathcal{T}^{\mu\nu}(\widetilde{k},\widetilde{k}',x_\perp)e_{\perp\nu} \, .
\end{split}
\end{equation}
The coordinates $x_\parallel$, $x_\perp$ and $\widetilde{x}$ are all independent variables since they are connected to the basis of orthogonal 4-vectors $e^\mu_\parallel$, $e^\mu_\perp$, and $\eta^\mu$. Therefore, both integrals over $dx_\parallel$ and $d\widetilde{x}$ would separate and produce a delta function. Choosing the integration over $dx_\parallel$ first yields $2\pi\delta[(\widetilde{k}-\widetilde{k}')/\mathcal{V}]$. Using this delta function to evaluate the $d\widetilde{k}'$ integral results in
\begin{equation}
 dp^\mu_\text{EM} = -\frac{2\pi x_\perp \mathcal{V}}{c}\int d\phi d\widetilde{k} d\widetilde{x} \,\mathcal{T}^{\mu\nu}(\widetilde{k},x_\perp)e_{\perp\nu} \, .
\end{equation}

In order to integrate the change of momentum, only a contraction of the kernel $\mathcal{T}^{\mu\nu}(\widetilde{k}, x_\perp)$ with the basis vector $e^\mu_\perp$ is necessary and the tensor structure simplifies. The term $\partial^\mu(I)$ is proportional to $e^\mu_\perp$ and the orthogonality conditions $e_\perp \cdot u = e_\perp \cdot \eta = 0$ apply, see \req{eq:orthogonality}. Using the kernel definition \req{eq:kernel} and substituting the electromagnetic and displacement tensors \reqs{eq:emtensor}{eq:displacementtensor} yield for the contraction
\begin{widetext}
 \begin{equation}
 \begin{split}
 \mathcal{T}^{\mu\nu}(\widetilde{k},x_\perp) e_{\perp\nu} &= - \frac{q^2\mu_0}{64\pi^2\mathcal{V}^2\Gamma^2}\left\{ i\widetilde{k}cI\left(1-\frac{1}{n^2\mathcal{V}^2}\right)(\eta^\mu - \Gamma u^\mu) \partial (I^{*})\cdot e_\perp - \partial^\mu(I)[\partial(I^{*})\cdot e_\perp](v \cdot u)\right.\\
 &+\left.\frac{1}{2}e^\mu_\perp\left[-\mathcal{V}^2\Gamma^2c^2\widetilde{k}^2 II^{*}n^2\left(1-\frac{1}{n^2\mathcal{V}^2}\right)^2+\partial(I)\cdot \partial(I^{*})(v\cdot u)\right]\right\}\,.
\end{split}
\end{equation}
\end{widetext}
When computing the integral over the polar angle $\phi$, the terms proportional to $\partial^\mu(I) \propto e^\mu_\perp$ and $e^\mu_\perp$ itself vanish. This is because the integration over $d\phi$ is an azimuthal average in the 2-plane orthogonal to $\eta^\mu$ and $u^\mu$. In that plane, $e_\perp^\mu$ is the outward unit four-vector normal to the cylinder surface and therefore rotates with $\phi$ and averages to zero. Therefore, the only remaining term is 
\begin{align} \label{eq:momentum_radiated}
\begin{split}
 dp^\mu_\text{EM} &= \frac{q^2\mu_0 x_\perp}{32 \pi \mathcal V\Gamma^2}\int_{-\infty}^{\infty}d\phi d\widetilde{k}d\widetilde{x} \, i\widetilde{k} I \\
 & \times\left(1-\frac{1}{n^2\mathcal V^2}\right)(\eta^\mu - \Gamma u^\mu)\partial (I^{*})\cdot e_\perp\,.
\end{split}
 \end{align}
Substituting $\partial^\mu(I^*)$ from \req{eq:partialmuI} and using $e_\perp^2 = -1$ makes the angular dependence trivial and the integral over $d\phi$ is equal to $2\pi$. The change in the field momentum is then
\begin{equation}
 dp^\mu_\text{EM} = -\frac{iq^2\mu_0 x_\perp}{16\mathcal{V}^2\Gamma^2}(\eta^\mu - \Gamma u^\mu) J(x_\perp)\,,
\end{equation}
where the remaining integral $J(x_\perp)$ reads
\begin{equation}
\begin{split}
 J(x_\perp) &\equiv \int_{-\infty}^{\infty} d\widetilde{k}d\widetilde{x}\, |\widetilde{k}|\widetilde{k}\left(1-\frac{1}{n^2\mathcal{V}^2}\right)\\
 &\times \sqrt{n^2\mathcal{V}^2-1}\, I \left(\frac{\partial I}{\partial \xi}\right)^*\,.
\end{split}
\end{equation}
This expression can be recast as an integral over positive wavenumbers
\begin{equation}\label{eq:Jintegral}
\begin{split}
 J(x_\perp) &= \int_{0}^{\infty} d\widetilde{k}d\widetilde{x}\, \widetilde{k}^2\left(1-\frac{1}{n^2\mathcal{V}^2}\right)\\
 &\times \sqrt{n^2\mathcal{V}^2-1}\, C(\widetilde{k})\,,
\end{split}
\end{equation}
where
\begin{equation}\label{eq:Ck}
 C(\widetilde{k}) \equiv I(\widetilde{k}) \left(\frac{\partial I(\widetilde{k})}{\partial \xi}\right)^* - I(-\widetilde{k}) \left(\frac{\partial I(-\widetilde{k})}{\partial \xi} \right)^*\,.
\end{equation}
For subluminal motion $n\mathcal{V} < 1$ the integral $I(\widetilde{k}) = H_0^{(1)}(\xi)$ for an arbitrary sign of the wavenumber $\widetilde{k}$ and the argument $\xi$ is purely imaginary [see solution \req{eq:Iintegralsol}]. The derivatives of the relevant Hankel functions read
\begin{equation}\label{eq:Hder}
 \frac{\partial H_0^{(1)}(\xi)}{\partial \xi} = - H_1^{(1)}(\xi)\,, \quad \frac{\partial H_0^{(2)}(\xi)}{\partial \xi} = - H_1^{(2)}(\xi)
\end{equation}
and since the wavenumber $\widetilde{k}$ enters into $\xi$ as an absolute value the factor 
\begin{equation}
 C(\widetilde{k}) = - H_0^{(1)} \left(H_1^{(1)}\right)^* + H_0^{(1)}\left(H_1^{(1)}\right)^* = 0
\end{equation}
in \req{eq:Jintegral} vanishes. 

For superluminal motion $n\mathcal{V} > 1$ the integral $I(\widetilde{k})$ is equal to [see \req{eq:Iintegralsol}]
\begin{equation}
 I(\widetilde{k}) = \Theta(\widetilde{k}) H_0^{(1)}(\xi) - \Theta(-\widetilde{k})H_0^{(2)}(\xi)
\end{equation}
and the argument $\xi$ is strictly real. Since the integral \req{eq:Jintegral} is over positive wavenumbers $\widetilde{k}$ the coefficient $C(\widetilde{k})$ [\req{eq:Ck}] is 
\begin{equation}
 C(\widetilde{k}) = H_0^{(1)} \left(\frac{\partial H_0^{(1)}}{\partial \xi}\right)^* - H_0^{(2)} \left(\frac{\partial H_0^{(2)}}{\partial \xi}\right)^*
\end{equation}
The functions $H_0^{(1)}$ and $H_0^{(2)}$ are for a real argument complex conjugate of each other so that
\begin{equation}
\begin{split}
 C(\widetilde{k}) &= H_0^{(1)} \frac{\partial H_0^{(2)}}{\partial \xi}- H_0^{(2)}\frac{\partial H_0^{(1)}}{\partial \xi}\\
 &= - H_0^{(1)} H_1^{(2)} + H_0^{(2)}H_1^{(1)} = - \frac{4i}{\pi\xi}\,,
\end{split}
\end{equation}
where the derivatives [\req{eq:Hder}] and Wronskian of the Hankel functions were evaluated. The change of the field momentum is then 
\begin{equation}\label{eq:dpEM}
\begin{split}
 dp^\mu_\text{EM} &= \frac{q^2 \mu_0}{4\pi \mathcal{V}\Gamma^2}(\Gamma u^\mu - \eta^\mu)\int_0^\infty d\widetilde{k} d\widetilde{x}\,\\
 &\times \widetilde{k} \left(1 - \frac{1}{n^2(\widetilde{k})\mathcal{V}^2}\right)\Theta[n(\widetilde{k})\mathcal{V}-1]\,.
\end{split}
\end{equation}
This radiated momentum change is equal and opposite to the momentum lost by the particle. Moreover, along the particle trajectory $x^\mu = z^\mu(\tau)$
\begin{equation}\label{eq:dxtilde}
d\widetilde{x} = \frac{d(x \cdot \eta)}{c} = \frac{u\cdot \eta}{c}d\tau = \Gamma c d\tau\,.
\end{equation}
Then the radiation friction force on the particle due to Cherenkov radiation is 
\begin{equation} \label{eq:particle_momentum}
\begin{split}
 \mathcal{F}_\mathrm{CRF}^\mu &= \frac{dp^\mu}{d\tau} = - \frac{dp^\mu_\text{EM}}{d\tau}= \frac{q^2\mu_0 c}{4\pi}\frac{\eta^\mu - \Gamma u^\mu}{\mathcal V \Gamma}\\
 &\times \int_0^\infty d\widetilde{k} \, \widetilde{k} \left(1-\frac{1}{n^2(\widetilde{k})\mathcal V^2}\right) \Theta[n(\widetilde{k})\mathcal V-1] \,. 
\end{split}
\end{equation}
Let us recall the antisymmetric tensor 
\begin{equation}
 R^{\mu\nu} \equiv \frac{1}{c^2}(\eta^\mu u^\nu - u^\mu \eta^\nu) \, ,
\end{equation}
then the covariant Cherenkov radiation friction force can be written as 
\begin{equation}
 \mathcal F^\mu_\text{CRF} = r_\text{CRF} R^{\mu\nu}u_\nu \, ,
\end{equation}
where the Cherenkov radiation friction coefficient is 
\begin{equation}
\begin{split}
 r_\text{CRF} = \frac{q^2\mu_0 c}{4\pi\mathcal V\Gamma}&\int_0^\infty d\widetilde{k} \,\widetilde{k}\left(1-\frac{1}{n^2(\widetilde{k})\mathcal V^2}\right)\\
 &\times\Theta[n(\widetilde{k})\mathcal V-1] \, .
\end{split}
\end{equation}
In order to compare directly with the Frank-Tamm formula, it is necessary to evaluate the stopping power. The radiated power (energy lost by the particle per proper time) can be expressed invariantly as a generalization of $\left.P_\text{rad}\right|_\text{M} = \left.dE/dt\right|_\text{M}$ for a laboratory observer in the medium rest frame. Such generalization reads
\begin{equation}\label{eq:prad}
\begin{split}
 P_\text{rad} &= \frac{d (\eta \cdot p)}{\Gamma d\tau} = -\frac{q^2\mu_0 c^3}{4\pi}\mathcal{V}\\
 &\times \int_0^\infty d\widetilde{k} \, \widetilde{k} \left(1-\frac{1}{n^2(\widetilde k)\mathcal V^2}\right) \Theta[n(\widetilde k)\mathcal V-1]\,.
\end{split}
\end{equation}
Assuming that for an observer in the medium rest frame, the particle travels a distance $dx$ in time $dt$ with speed $\left. \mathcal{V}\right|_\text{M} = \beta$ 
\begin{equation}
\begin{split}
 &\left.\frac{dE}{dx}\right|_\text{M} = \left. \frac{dE}{\beta c dt}\right|_\text{M} = \left.\frac{P_\text{rad}}{\beta c}\right|_\text{M} = -\frac{q^2\mu_0 c^2}{4\pi}\\
 &\times \int_0^\infty d k^0 \, k^0 \left(1-\frac{1}{n^2(k^0)\mathcal \beta^2}\right) \Theta[n(k^0)\beta-1] 
\end{split}
\end{equation}
which coincides with the Frank-Tamm formula \req{eq:Frank-Tamm} with angular frequency $\omega = c k^0$. 

The prescription for the change of the four-momentum of the electromagnetic field in \req{eq:dpEM} allows us to derive the emission spectrum of Cherenkov radiation. Considering that the invariant energy of the emitted photons is $\eta \cdot p_\text{EM}$ the number of emitted photons with a given wavenumber can be calculated by dividing this energy by an energy of a photon quantum 
\begin{equation}
dN = \frac{\eta \cdot dp_\text{EM}}{\hbar c \widetilde{k}} 
\end{equation}
and using \req{eq:dpEM} we obtain
\begin{equation}
\frac{dN}{d\widetilde{x}d\widetilde{k}} = \frac{q^2 \mu_0 c}{4\pi \hbar} \mathcal{V}\left(1-\frac{1}{n^2(\widetilde k)\mathcal V^2}\right) \Theta[n(\widetilde k)\mathcal V-1]\,.
\end{equation}
Finally, after integrating the coordinate $d\widetilde{x}$ along the particle trajectory [\req{eq:dxtilde}] the emission spectrum reads
\begin{equation}
\begin{split}
\frac{dN}{d\widetilde{k}} &=\frac{q^2 \mu_0 c^2}{4\pi \hbar} \mathcal{V} \Gamma\\
&\times \int d\tau\, \left(1-\frac{1}{n^2(\widetilde k)\mathcal V^2}\right) \Theta[n(\widetilde k)\mathcal V-1]\,.
\end{split}
\end{equation}
For superluminal uniform motion the integrand is nonzero and constant, therefore in the steady state solution considered here (ignoring friction), the number of photons grows linearly with proper time. Then the emission spectrum is determined by the $n(\widetilde{k})$ dependence, which is a property of the specific material medium. 

For water in the 200 - 800 nm transparency region, $n(\widetilde{k})$ is approximately constant and the spectrum is flat. We applied the results above to obtain estimates of the magnitude of the Cherenkov radiation friction force experienced by an ultrarelativistic electron in water, presented in~\reqs{aCRF1}{aCRF2}. Consideration of the more interesting application domain, the relativistic quark-gluon plasma, is deferred to topic-dedicated work, see additional comments in the following section.

\section{Conclusions and discussion}\label{sec:discussion}
We reformulated in a covariant manner the Cherenkov radiation emission by
a charged particle passing through a dielectric medium with homogeneous and isotropic index of refraction satisfying $n(\widetilde k) > 0$, $\text{Im}[n(\widetilde k)] = 0$ (no absorption), $dn(\widetilde k)/d\widetilde k>0$ (regular dispersion). We employed covariant constitutive relations to derive the electromagnetic field and the displacement field of a charged particle traveling through the medium at a constant relative speed exceeding the local speed of light. This generalizes the Cherenkov condition in terms of the invariant index of refraction as well as the invariant relative velocity between the particle and the medium. 

We then obtained Cherenkov radiation friction by evaluating the change of four-momentum emitted via radiation per increment of particle proper time which we consider to be opposite to the Cherenkov radiation friction force. This is doable because the Cherenkov radiation friction force derived here is orthogonal to the particle four-velocity, and hence preserves the magnitude of relativistic momentum $p^2 = m^2c^2$. In fact, the derived Cherenkov radiative friction force has the same tensor structure as a covariant material friction force \cite{dunkel2009relativistic,Formanek:2020zwc}. This orthogonality is only possible since the Cherenkov force in our covariant formulation requires the use of an additional four-vector, the medium four-velocity. 

Cherenkov radiation requires the existence of a frame of reference describing the medium in which the particle moves. In prior works Cherenkov radiation was treated in the rest frame of the medium. However, the covariant formulation was previously explored in more exotic contexts, such as that of tachyonic radiation~\cite{Jones:1972cfq,Perepelitsa:2015vxa}. Our formulation is fundamentally different as it investigates in-medium superluminal charged particle motion with speed higher than that of in-medium electromagnetic wave propagation unlike radiation of hypothetical superluminal particles in vacuum. Covariant formulation was also implemented in description of radiation appearing in modified Maxwell theory in vacuum with a symmetry breaking term mimicking non-inertial particle dynamics \cite{Lehnert:2004hq}.

This work focused on an idealized dielectric without any absorption (see also discussions in Appendices \ref{app:contour_int} and \ref{app:proper_time_int}). With a non-zero imaginary part of the index of refraction, the radiated field could have a non-zero momentum change even in the subluminal motion case, accounting for the field absorption in the medium. Since the goal here was to compare our result with the Frank-Tamm formula in a non-absorptive dielectric, we will leave the consideration of dissipative environments to future work. 

Our covariant generalization can impact many diverse physical environments involving the motion of relativistic charged particles. Cherenkov radiation can also be an important source of radiation in relativistic laser-matter interactions \cite{DiPiazza:2011tq,Burton:2014wsa,Blackburn:2019rfv,Gonoskov:2021hwf,Fedotov:2022ely} and warrants further scrutiny. Cherenkov radiation is not equivalent to the usual bremsstrahlung consideration of soft photon production in charged particle collisions. Therefore, one should expect an excess of soft photons when conditions for emission of Cherenkov radiation are satisfied such as in relativistic cosmic and laboratory environments. Indeed, such an excess beyond collisional QED prediction accompanies many if not all high energy laboratory collision processes. The measurement of this effect is not easy, yet it has been reported for several individual hadron collisions; see Ref. \cite{belogianni2002observation} and references therein. 

One could wonder where the dielectric medium in such collisions is, so that the colliding relativistic and electrically charged quarks could emit Cherenkov radiation: In recent years signatures of equilibrated, relatively small volume quark-gluon plasma formation have been reported in $pp$ collisions at the LHC. At lower collision energies, the deconfinement volume is smaller and thermal equilibration may not be achieved due to a shorter lifespan. However, we can expect the presence of a dielectric deconfined medium and thus the ability of abundantly present relativistic charged particles to produce Cherenkov radiation. We believe that our study of the covariant Cherenkov radiation creates the tools to study related radiative processes, the resulting spectrum of soft photons provides an opportunity for diagnosis of the electromagnetic properties of the deconfined phase of matter.

Beyond potential use in study of soft photon production, we are very interested in generalizing our results for the case of an accelerated charged particle within a medium, allowing for the combined Cherenkov radiation due to superluminal motion and Larmor radiation due to acceleration. The presence of a medium reference frame four-velocity could help resolve the inconsistencies that plague the theory of radiation friction in an ideal vacuum, which we have mentioned. 

We believe that the issues associated with Larmor radiation could be addressed in our approach. Normally, Larmor radiation is studied  using the ideal vacuum state, which is an idealization: The Universe emerged from a dense primordial plasma phase and therefore a perfect vacuum never existed. The Larmor radiation emission could then be derived with reference to the cosmological cosmic microwave background frame of reference \cite{Durrer:2015lza,ellis2012relativistic}, which our formulation allows. Moreover, we hope to explore in the covariant manner Cherenkov radiation emitted in a vacuum sufficiently polarized by a strong external field \cite{Macleod:2018zcb}.

\appendix

\section{Coordinate system}\label{sec:coords}
\begin{figure}
 \centering
 \includegraphics[width=1.0\linewidth]{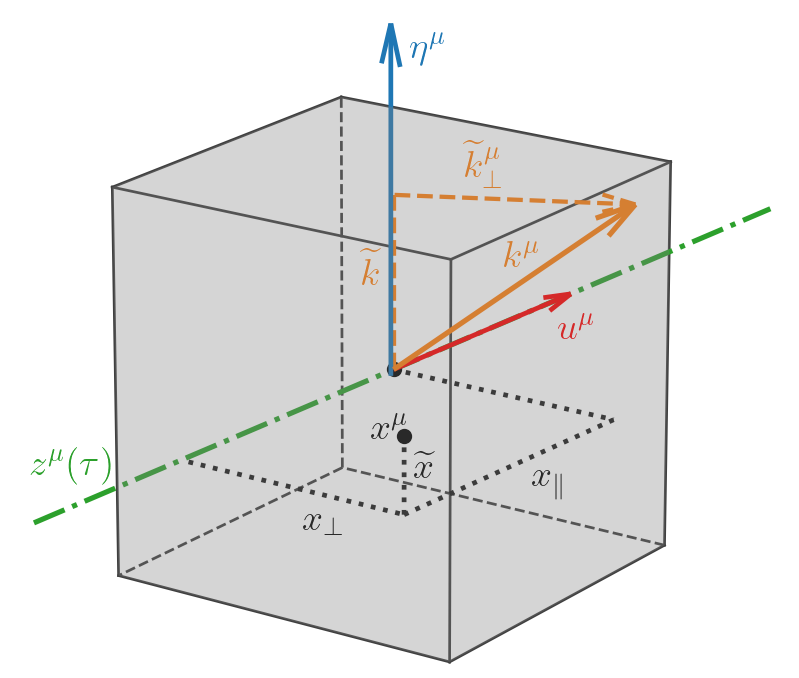}
 \caption{\label{fig:illustration}Schematic 2D representation of the 4D problem. Within a material medium with a constant four-velocity $\eta^\mu$ a charged particle is moving along a trajectory $z^\mu(\tau)$ (dash-dotted line). The four-velocity of the particle is a constant $u^\mu$. If the particle is superluminal within this medium it emits radiation represented by a spectral component with a four-wavevector $k^\mu$ which can be decomposed to an invariant wavenumber $\widetilde{k}$ (dashed line) and perpendicular four-vector $\widetilde{k}_\perp^\mu$ (dashed arrow). A general position four-vector $x^\mu$ can be also decomposed into components (dotted lines): (a) parallel to medium four-velocity $\widetilde{x}$ [\req{eq:x_tilde}], and in a subspace orthogonal to medium four-velocity into components (b) parallel to particle motion $x_\parallel$ [\req{eq:x_parallel}] and (c) transverse to four-velocity $x_\perp$ [\req{eq:x_perp}].}
 
\end{figure}

The introduction of the additional medium four-velocity $\eta^\mu$ allows us to build a coordinate system. Let us define a projection tensor $P^{\mu\nu}$ on a subspace orthogonal to the medium four-velocity $\eta^\mu$ as
\begin{equation}\label{eq:projector}
 P^{\mu\nu} \equiv g^{\mu\nu} - \frac{\eta^\mu\eta^\nu}{c^2} \, .
\end{equation}
An arbitrary four-vector $w^\mu$ can be decomposed into components parallel and perpendicular to $\eta^\mu$ using this projector as
\begin{equation}\label{eq:etadecompose}
\begin{split}
 w^\mu = \widetilde{w} &\frac{\eta^\mu}{c} + \widetilde{w}_\perp^\mu\,, \\
 \widetilde{w} \equiv \frac{w\cdot \eta}{c}\,, &\quad \widetilde{w}_\perp^\mu \equiv P^{\mu\nu}w_\nu\,.
\end{split}
\end{equation}
This decomposition is used here to separate the components of the wave four-vector $k^\mu$ and the position four-vector $x^\mu$ to evaluate integrals. For example, following this convention, the coordinate along the four-velocity of the medium is
\begin{equation}\label{eq:x_tilde}
 \widetilde{x} \equiv \frac{x \cdot \eta}{c}\,. 
\end{equation}

On the 3D subspace perpendicular to $\eta^\mu$ the Lorentz invariant coordinates can be defined with the meaning of coordinates parallel and transverse to the motion of the particle characterized by its instantaneous four-velocity $u^\mu$ (see Fig.~\ref{fig:illustration})
\begin{align}
 x_\parallel &\equiv - \frac{x\cdot P\cdot u}{\mathcal V \Gamma c} = -\frac{\widetilde{x}_\perp \cdot u}{\mathcal{V}\Gamma c}\,,\label{eq:x_parallel}\\
 x_\perp &\equiv \sqrt{ -x \cdot P \cdot x - x_\parallel^2} = \sqrt{-\widetilde{x}_\perp^2-\frac{(\widetilde{x}_\perp\cdot u)^2}{\mathcal{V}^2\Gamma^2c^2}}\label{eq:x_perp}\, .
\end{align}
For an observer in the medium rest frame observing a particle moving along the $z$ direction, they reduce to 
\begin{equation}
 \left. x_\parallel\right|_\text{M} = z\,, \quad \left. x_\perp \right|_\text{M} = \sqrt{x^2+y^2}\,.
\end{equation}
The corresponding basis vectors are 
\begin{align}
 e_\parallel^\mu &\equiv \partial^\mu x_\parallel = - \frac{P^{\mu\nu}u_\nu}{\mathcal{V}\Gamma c}\,,\label{eq:e_parallel}\\
 e_\perp^\mu &\equiv \partial^\mu x_\perp = -\frac{1}{x_\perp}\left(P^{\mu\nu}x_\nu + \frac{x\cdot P\cdot u}{\mathcal V^2\Gamma^2 c^2}P^{\mu\nu}u_\nu\right)\label{eq:e_perp}\,.
\end{align}
In the rest frame of the medium, the four-vector $e^\mu_\perp$ points inward toward the particle trajectory and the four-vector $e^\mu_\parallel$ points against the particle velocity. 

Together with $\eta^\mu$, which is proportional to the basis vector $\partial^\mu \widetilde{x}$, these four-vectors are mutually orthogonal
\begin{equation}\label{eq:orthogonality}
 \eta \cdot e_\parallel = \eta \cdot e_\perp = e_\parallel \cdot e_\perp = 0\,,
\end{equation}
and have magnitudes $\eta^2 = c^2$, $e_\parallel^2 = e_\perp^2 = -1$. Finally, by design $e_\perp^\mu$ is orthogonal to particle four-velocity and $e_\parallel^\mu$ has a non-zero overlap with particle four-velocity
\begin{equation}
 u \cdot e_\perp = 0\,, \quad u \cdot e_\parallel = \mathcal{V} \Gamma c\,.
\end{equation}
For completeness, it was already established that $u^2 = c^2$ and $u \cdot \eta = \Gamma c^2$. Therefore, $u^\mu$ is a linear combination of $\eta^\mu$ and $e^\mu_\parallel$, namely $u^\mu = \Gamma \eta^\mu - \mathcal{V}\Gamma c e^\mu_\parallel$. The case of $u^\mu$ and $\eta^\mu$ being completely linearly dependent (and equal because they are normalized to $c^2$) corresponds to a trivial case of a particle at rest in the material medium ($\mathcal{V} = 0$, $\Gamma = 1$). Finally, an arbitrary four-vector $w^\mu$ can be decomposed as
\begin{equation}\label{eq:decomposetot}
 w^\mu = \widetilde{w}\frac{\eta^\mu}{c} - w_\parallel e^\mu_\parallel - w_\perp e^\mu_\perp\,,
\end{equation}
where $w_\parallel = w\cdot e_\parallel$ and $w_\perp = w \cdot e_\perp$. This further develops the decomposition from \req{eq:etadecompose} in the 3D subspace transverse to $\eta^\mu$.

\section{Calculation of momentum space propagator and choice of gauge}\label{app:inverse}
Defining the tensor on the left-hand side of \req{eq:Av} as $\Lambda^\mu _\nu$ while working in the gauge $k \cdot A = 0$
\begin{equation}
\begin{split}
 \Lambda^\mu_{\,\,\,\nu} &\equiv \left[k^2+(n^2-1)\widetilde{k}^2\right] \delta^\mu_\nu \\ 
 &+(n^2-1)\left(k^2\eta^\mu \eta_\nu - \widetilde{k} ck^\mu\eta_\nu \right)/c^2\,.
\end{split}
\end{equation}
Then its inverse can be written in terms of a cofactor matrix $C^\mu_{\,\,\,\nu}$ [see Eqs. (2.2.13) and (2.2.14) in Ref.~\cite{Melrose:qu2008}]
\begin{equation}
 (\Lambda^{-1})^\mu_{\,\,\,\nu} = \frac{C^\mu_{\,\,\,\nu}}{\text{det}\Lambda} \, .
\end{equation}
First, we define $\Lambda^{(n)}$ as the trace of the nth power of $\Lambda^\mu_{\,\,\,\nu}$
\begin{equation}
 \Lambda^{(1)} = \Lambda^\mu_{\,\,\,\mu}\,, \quad \Lambda^{(2)} = \Lambda^\mu_{\,\,\,\nu}\Lambda^\nu_{\,\,\,\mu}\,,\quad \ldots
\end{equation}
Then the formulae for the determinant and the cofactor matrix are
\begin{equation} \label{eq:determinant}
\begin{split}
 \det\Lambda &= \tfrac{1}{24}\big[
 (\Lambda^{(1)})^4 
 + 8\Lambda^{(1)}\Lambda^{(3)} 
 + 3(\Lambda^{(2)})^2\\
 &- 6(\Lambda^{(1)})^2\Lambda^{(2)}
 - 6\Lambda^{(4)}
 \big]\,,
\end{split}
\end{equation}
\begin{equation} \label{eq:cofactor}
C^\mu_{\,\,\,\nu} = 
C_1\delta^\mu_{\,\,\,\nu} 
+ C_2 \Lambda^\mu_{\,\,\,\nu} 
+ C_3 (\Lambda^2)^\mu_{\,\,\,\nu} 
- (\Lambda^3)^\mu_{\,\,\,\nu} \,,
\end{equation}
where the coefficients are given by
\begin{align}
 C_1 &= \frac{1}{6}\left[(\Lambda^{(1)})^3-3\Lambda^{(1)}\Lambda^{(2)}+2\Lambda^{(3)}\right]\,, \\\
 C_2 &= -\frac{1}{2}\left[(\Lambda^{(1)})^2-\Lambda^{(2)}\right]\,, \\
 C_3 &= \Lambda^{(1)} \,.
\end{align} 
The propagator in the momentum space can be calculated as 
\begin{equation}
\begin{split}
 D^{\mu\nu}(k) &= - (\Lambda^{\mu\nu})^{-1} =\\
 &-\frac{k^2 P_n^{\mu\nu} + (1-1/n^2)\widetilde{k} k^\mu\eta^\nu/c}{k^2\left[k^2 + (n^2-1) \widetilde{k}^2\right]}\\
 &\equiv D^{\mu\nu}_1(k) + D^{\mu\nu}_2(k)\, ,
\end{split}
\end{equation}
where the tensor structure of the term $D_1^{\mu\nu}$ is contained in $P_n^{\mu\nu}$ defined in \req{eq:Pmn}. The second term in the propagator nominator will not contribute to the electromagnetic field tensor. We can see this through an explicit computation. The potential is calculated from the propagator as \req{eq:potential}
\begin{equation}
 A^\mu(k) = \mu_0 D^{\mu\nu}(k) j_\nu(k) \,.
\end{equation}
The electromagnetic field tensor then reads
\begin{equation}
\begin{split}
 F^{\mu\nu}(k) &= -i[k^\mu A^\nu(k) - k^\nu A^\mu(k)] \, \\
 &= 2i\mu_0\frac{k^{[\mu}j^{\nu]} - (1-1/n^2)(\eta\cdot j)k^{[\mu}\eta^{\nu]}}{k^2 + (n^2-1) \widetilde{k}^2}\\
 &+ 2i\mu_0\frac{(1-1/n^2)\widetilde{k}(\eta\cdot j) k^{[\mu}k^{\nu]}/c}{k^2[k^2 + (n^2-1) \widetilde{k}^2]} \, ,
\end{split}
\end{equation}
where the square brackets indicate the antisymmetric combination of the enclosed indices \req{eq:antisym}. The second term will vanish since $k^\mu k^\nu$ is a symmetric combination. To understand why $D_2^{\mu\nu}$ does not contribute to $F^{\mu\nu}$, we can write out the corresponding potential in the position space as
\begin{equation}
 A_2^\mu(x) = \mu_0 \int d^4x' D_2^{\mu\nu}(x-x') j_\nu(x') =\partial^\mu \lambda \, ,
\end{equation}
where we have defined $\lambda$ as
\begin{equation}
\begin{split}
 \lambda(x) &\equiv - i \mu_0 \int d^4x' \int \frac{d^4 k}{(2\pi)^4} \\
 &\times \frac{(1-1/n^2)\widetilde{k}[\eta \cdot j(x')]/c}{k^2[k^2+(n^2-1)\widetilde{k}^2]} e^{-ik\cdot(x-x')} \, .
\end{split}
\end{equation}
Since $A_2^\mu$ can be expressed as a four-gradient of a scalar it is effectively a gauge transformation. That is why the gauge-invariant tensor $F^{\mu\nu}$ does not depend on $D_2^{\mu\nu}$.

\section{Contour integral for the position space propagator}\label{app:contour_int}
The position-space form of $D^{\mu\nu}(k)$ is given by an inverse Fourier transformation
\begin{equation}
 D^{\mu\nu}(x-x') = -\int \frac{d^4k}{(2\pi)^4} \frac{P_n^{\mu\nu}(\widetilde k) e^{-ik\cdot (x-x')}}{k^2 + [n^2(\widetilde k)-1]\widetilde{k}^2} \,.
\end{equation}
We can decompose $k^\mu$ into a component $\widetilde{k}$ parallel to $\eta^\mu$ and components $\widetilde{k}_\perp^\mu$ perpendicular to $\eta^\mu$ according to \req{eq:etadecompose}. Then, the propagator becomes
\begin{equation}
\begin{split}
 D^{\mu\nu}(x-x') &= -\int \frac{d\widetilde{k}}{(2\pi)^4} e^{-i \widetilde{k} \eta \cdot (x-x')/c}\\
 &\times P_n^{\mu\nu}(\widetilde{k}) I_\perp (\widetilde{k},x-x')\,,
\end{split}
\end{equation}
where the integral over the transverse momentum is
\begin{equation}\label{eq:Iperp}
 I_\perp(\widetilde{k},x-x') = \int d^3\widetilde{k}_\perp \frac{e^{-i\widetilde{k}_\perp \cdot (x-x')}}{\widetilde{k}_\perp^2 + n^2(\widetilde{k}) \widetilde{k}^2}\,.
\end{equation}
Since this integral is an invariant quantity, it does not matter in which frame it is calculated. We choose an observer situated in the rest frame of the medium (denoted by subscript M) for which
\begin{align}
 \left. \eta^\mu \right|_\text{M} &= (c,0,0,0) \, ,\\
 \left.\widetilde{k}\right|_\text{M}= k^0, &\quad \left. \widetilde{k}_\perp^\mu \right|_\text{M} = (0,\boldsymbol k_\perp) \, .
\end{align}
Then the integral in \req{eq:Iperp} has a form
\begin{equation}
 \left. I_\perp(\widetilde{k}, x - x')\right|_\text{M} = \int d^3 k_\perp \frac{e^{i \boldsymbol{k}_\perp\cdot (\boldsymbol{x}-\boldsymbol{x}')}}{-\kappa^2 + n^2(k^0) (k^0)^2}\,,
\end{equation}
where $\kappa \equiv |\boldsymbol{k}_\perp|$. Except for the function $n(k^0)$ in the denominator, this integral is equivalent to the integral solved in Chapter 43.2 of book~\cite{schwinger2000classical}. Using cylindrical coordinates 
\begin{equation}
 I_\perp|_\text{M} = \int d\kappa \kappa^2 d(\cos\theta) d\phi \frac{e^{i\kappa R\cos\theta}}{-\kappa^2 + n^2(k^0) (k^0)^2}\,,
\end{equation}
where 
\begin{equation}
 R \equiv |\boldsymbol x - \boldsymbol x'| \,.
\end{equation}
Carrying out the angular integrals yields
\begin{equation}
\begin{split}\label{eq:complexintegral}
 \left. I_\perp\right|_\text{M} &= \frac{2 \pi i}{R} \int_0^\infty d\kappa \kappa \frac{e^{i\kappa R}-e^{-i\kappa R}}{\kappa^2 - n^2(k^0) (k^0)^2}\\
 &=\frac{2\pi i}{R} \int_{-\infty}^\infty d\kappa \frac{\kappa e^{i\kappa R}}{\kappa^2 - n^2(k^0) (k^0)^2} \,,
\end{split}
\end{equation}
where the two terms in the integral over $\kappa$ were combined. Usually, $k^0$ would now be analytically continued as $k^0 + i \epsilon$, with $\epsilon > 0$ so that the retarded solution is selected. This can also be done here, assuming that $n(k^0)$ is an analytic function in the upper half of the complex $k^0$ plane. The poles of the denominator are 
\begin{equation}
 \kappa = \pm n(k^0+i\epsilon)(k^0 + i \varepsilon)\,.
\end{equation}
If we perform a Taylor expansion of the index of refraction, the poles are
\begin{equation}\label{eq:poles}
\begin{split}
 \kappa &\approx \pm \{n(k^0) k^0 + i\epsilon[n(k^0)+n'(k^0) k^0]\}\,.
\end{split}
\end{equation}
In order for the integral to converge, the integration contour has to be closed in the upper half plane. Assuming a regular dielectric medium such that $n(k^0)> 0$ with no imaginary part, the pole with a plus sign applies to both positive and negative wavenumbers. This is because $n(k^0)$ is an even function of wavenumber and its derivative is odd. As a side note, typical dielectric mediums have a small but non-zero absorption properties manifested through imaginary part of the index of refraction. In such a case, the locations of the poles for the propagator in momentum space \req{eq:poles} shift to
\begin{equation}
\begin{split}
 \kappa &= \pm \{\text{Re}[n(k^0)] k^0 + i\text{Im}[n(k^0)]k^0 \\
 &+ i\epsilon[n(k^0)+n'(k^0) k^0]\}\,.
\end{split}
\end{equation}
Now, the imaginary part of the index of refraction alone is sufficient to regularize the integral \req{eq:complexintegral}. For both positive and negative wavenumbers the pole with the plus sign moves above the real axis due to $\text{Im}[n(-k^0)] = - \text{Im}[n(k^0)]$. 

Returning to the non-absorptive dielectric and using the positive pole from \req{eq:poles} and the residue theorem, the integral over perpendicular momenta is in the limit $\epsilon \rightarrow 0^+$
\begin{equation}
\begin{split}
 \left. I_\perp\right|_\text{M} &= \frac{2\pi i}{R}\int_{-\infty}^{\infty}d\kappa \frac{\kappa e^{i\kappa R}}{\kappa^2 - n^2(k^0)(k^0)^2}\\
 &=-\frac{2\pi^2}{R} e^{in(k^0) k^0 R}\,.
\end{split}
\end{equation}
The result of this integration is written in terms of quantities in the medium rest frame. However, the original integral was Lorentz invariant, so this result holds in any frame. An invariant generalization of the quantities $k^0$ and $R$ is
\begin{align}
 k^0 &\rightarrow \widetilde{k}\, , \\
 R &\rightarrow \sqrt{-(\widetilde{x}_\perp-\widetilde{x}'_\perp)^2} \,,
\end{align}
where $\widetilde{x}^\mu_\perp - \widetilde{x}'^\mu_\perp = P^{\mu\nu}(x-x')_\nu$ is the difference in the transverse coordinates according to the decomposition in \req{eq:etadecompose}. Then the result for the propagator in any frame reads
\begin{equation} \label{eq:propagator2}
\begin{split}
 D^{\mu\nu}(x-x') &= \frac{1}{8\pi^2}\int_{-\infty}^{\infty}d\widetilde{k} P_n^{\mu\nu}(\widetilde k) \\
 &\times \frac{e^{-i\widetilde{k}\left[\widetilde{x}-\widetilde{x}' - n(\widetilde k)\sqrt{-(\widetilde{x}_\perp-\widetilde{x}'_\perp)^2}\right]}}{\sqrt{-(\widetilde{x}_\perp-\widetilde{x}'_\perp)^2}} \,,
\end{split}
\end{equation}
where $\widetilde{x}-\widetilde{x}' = (x-x')\cdot \eta /c$ is the difference in the coordinate along the medium four-velocity according to the decomposition in \req{eq:etadecompose}.
\section{Proper time integral for $A^\mu$}\label{app:proper_time_int}
With the charge particle current density prescribed by its trajectory $z^\mu(\tau)$ and four-velocity $u^\mu(\tau)$ \cite{Rafelski:2017hyt}
\begin{equation}
 j^\mu(x) = qc \int_{-\infty}^\infty d\tau \, u^\mu(\tau) \delta^4[x-z(\tau)] \,,
\end{equation}
the four-potential is given by \req{eq:potential2}
\begin{equation} \label{eq:potential_1}
\begin{split}
 A^\mu(x) &= \frac{\mu_0 q c}{8\pi^2}\int_{-\infty}^\infty d\widetilde{k} \, P_n^{\mu\nu}(\widetilde{k})
 \int_{-\infty}^\infty d\tau \, u_\nu(\tau)\\
 &\times \frac{e^{-i\widetilde{k}\left(\widetilde{x} - \widetilde{z}(\tau) - n(\widetilde{k})\sqrt{-[\widetilde{x}_\perp - \widetilde{z}_\perp(\tau)]^2}\right)}}{\sqrt{-[\widetilde{x}_\perp - \widetilde{z}_\perp(\tau)]^2}} \,.
\end{split}
\end{equation}
The scalar appearing under the square root is
\begin{equation}\label{eq:projection}
\begin{split}
 -&(\widetilde{x}_\perp - \widetilde{z}_\perp)^2 = -(x-z) \cdot P \cdot (x-z) \\
 &= -x \cdot P \cdot x + 2x \cdot P\cdot z - z \cdot P\cdot z \, .
\end{split}
\end{equation}
Adding and subtracting $x_\parallel^2$ and decomposing the observer position and the particle trajectory according to \req{eq:decomposetot}, the expression in \req{eq:projection} can be recast as
\begin{equation}
 -[\widetilde{x}_\perp - \widetilde{z}_\perp(\tau)]^2 = [x_\perp - z_\perp(\tau)]^2 + [x_\parallel-z_\parallel(\tau)]^2\,,\\
\end{equation}
for any general trajectory $z^\mu(\tau)$. In this Appendix, the proper time integral in \req{eq:potential_1} will be computed for a uniformly moving particle, whose trajectory is a linear function of the proper time
\begin{equation}
 z^\mu(\tau) = u^\mu\tau\,,
\end{equation}
where $u^\mu$ is a constant four-vector, and $z^\mu(0)=0$. In this case, the basis vectors $e^\mu_\perp$ and $e^\mu_\parallel$ are constant, and the trajectory components are
\begin{equation}
\begin{split}
\widetilde{z} &= z \cdot \frac{\eta}{c} = \Gamma c \tau\,,\\
\quad z_\perp = z\cdot e_\perp &= 0\,, \quad z_\parallel = z \cdot e_\perp = \mathcal{V}\Gamma c \tau\,.
\end{split}
\end{equation}
Therefore, the expression in \req{eq:projection} is
\begin{equation}
-[\widetilde{x}_\perp - \widetilde{z}_\perp(\tau)]^2 = x_\perp^2 +(x_\parallel-\mathcal{V}\Gamma c\tau)^2\,.
\end{equation}
Now we perform a change of variables
\begin{equation}
 \zeta \equiv \frac{x_\parallel - \mathcal V\Gamma c\tau}{x_\perp} \, ,
\end{equation}
under which the expression in the exponential part of the proper time integral can be written as
\begin{equation}
\begin{split}
 &\widetilde{x}-\widetilde{z}(\tau) - n(\widetilde k)\sqrt{-[\widetilde{x}_\perp - \widetilde{z}_\perp(\tau)]^2} \\
 &= \widetilde{x} - \frac{x_\parallel - x_\perp \zeta}{\mathcal{V}} - n(\widetilde{k})x_\perp\sqrt{1+\zeta^2} \\
 &= \widetilde{x} - \frac{x_\parallel}{\mathcal{V}} + \frac{x_\perp}{\mathcal{V}}\left[\zeta-n(\widetilde{k})\mathcal{V}\sqrt{1+\zeta^2}\right] \,.
\end{split}
\end{equation}
The first two terms in the exponent are a constant phase factor, so that the four-potential in~\req{eq:potential_1} can ultimately be written as
\begin{equation} \label{eq:app_potential}
\begin{split}
 A^\mu(x) = \frac{i\mu_0 q}{8\pi\mathcal V\Gamma}&\int_{-\infty}^\infty d\widetilde{k} \, P_n^{\mu\nu}(\widetilde k)u_\nu\\
 &\times e^{-i\widetilde{k}(\widetilde{x}-x_\parallel/\mathcal{V})} I(\widetilde{k},x_\perp)\,,
\end{split}
\end{equation}
and the following integral needs to be computed
\begin{equation}
 I(\widetilde{k},x_\perp) \equiv
 \frac{1}{i\pi} \int_{-\infty}^{\infty} d\zeta \frac{e^{-i\widetilde{k} \frac{x_\perp}{\mathcal V}\left(\zeta-n\mathcal{V}\sqrt{1+\zeta^2}\right)}}{\sqrt{1+\zeta^2}} \,.
\end{equation}
The $1/i\pi$ prefactor was chosen for later convenience. 

The unitless scalar integral $I(\widetilde{k})$ can be further simplified by changing the integration variable to $\zeta \equiv \sinh \varphi$ so that 
\begin{equation}\label{eq:Iintegral}
 I(\widetilde{k},x_\perp) = \frac{1}{i\pi}\int_{-\infty}^{\infty} d\varphi \, e^{-i\widetilde{k} \frac{x_\perp}{\mathcal V}(\sinh \varphi - n \mathcal{V} \cosh \varphi )} \, .
\end{equation}
The integration of~\req{eq:Iintegral} has different outcomes depending on the value of $n\mathcal V$, so we consider the cases of $n\mathcal V<1$ and $n\mathcal V>1$ separately. Note that in case of an absorptive medium with a non-zero imaginary part of $n$ such sharp Cherenkov regimes $n\mathcal{V} > 1$ and $n\mathcal{V} < 1$ can no longer be distinguished.

\subsection{Case $n\mathcal V < 1$}
In this case, the rapidity-like variable $y_1$ can be introduced such that
\begin{align}
 \cosh y_1 &= \frac{1}{\sqrt{1-n^2 \mathcal V^2}}\,, \\
 \tanh y_1 &= n \mathcal V\, .
\end{align}
The argument of the exponential in the integrand of \req{eq:Iintegral} can then be simplified to
\begin{equation}
\begin{split}
 -\frac{i\widetilde{k} x_\perp}{\mathcal V \cosh y_1}&(\cosh y_1 \sinh\varphi - \sinh y_1 \cosh\varphi) \\
 &= -i\widetilde{k} x_\perp \frac{\sqrt{1-n^2\mathcal V^2}}{\mathcal V}\sinh \left(\varphi-y_1\right)\,.
\end{split}
\end{equation}
Let us define a quantity
\begin{equation}
 \sigma_1 \equiv \frac{|\widetilde{k}|x_\perp}{\mathcal V}\sqrt{1-n^2 \mathcal V^2}\,,
\end{equation}
which is a real positive number for $n\mathcal V<1$. For positive wavenumbers $\widetilde{k}>0$, the integral \req{eq:Iintegral} is
\begin{equation} \label{eq:integral2}
 I_1(\widetilde{k} > 0) = \frac{1}{i\pi}\int_{-\infty}^{\infty} d\varphi \, e^{-i\sigma_1 \sinh (\varphi-y_1)},
\end{equation}
while for negative wavenumbers this integral is 
\begin{equation} \label{eq:integral3}
 I_1(\widetilde{k} < 0) = \frac{1}{i\pi}\int_{-\infty}^{\infty} d\varphi \, e^{i\sigma_1 \sinh (\varphi-y_1)} \,.
\end{equation}
The result of these integrations is the same in either case and can be evaluated using the integral representation of the Macdonald function of zeroth order $K_0(x)$~\cite{Abramhowitz:po1984}
\begin{equation}
 K_0(x) = \frac{1}{2}\int_{-\infty}^{\infty} dt \, e^{-ix\sinh t}\,,
\end{equation}
which holds for $x$ real and positive. Then~\req{eq:integral2} and~\req{eq:integral3} can be expressed as 
\begin{equation}
 I_1(\widetilde{k},x_\perp) = \frac{2}{i\pi} K_0\left(\sigma_1\right)
\end{equation}
for arbitrary $\widetilde{k}$. This result can be written in terms of the zeroth order Hankel function of the first kind using the relation~\cite{Abramhowitz:po1984}
\begin{equation}
 K_0(x) = \frac{i\pi}{2}H_0^{(1)}(ix) \, ,
\end{equation}
which again holds for $x$ real and positive. Finally, the integral \req{eq:Iintegral} is
\begin{equation}\label{eq:I1}
 I_1(\widetilde{k},x_\perp) = H_0^{(1)}(i\sigma_1)
\end{equation}
for subluminal motion $n\mathcal V<1$.

\subsection{Case $n\mathcal V>1$}
Starting from~\req{eq:Iintegral}, the rapidity-like variable $y_2$ can be defined in this regime as
\begin{align}
 \tanh y_2 &= \frac{1}{n\mathcal V}\,,\\
 \sinh y_2 &= \frac{1}{\sqrt{n^2\mathcal V^2-1}}\,,
\end{align}
so that the square root remains real. The argument of the exponential in the integrand of~\req{eq:Iintegral} can be simplified as 
\begin{equation}
\begin{split}
 -\frac{i\widetilde{k} x_\perp}{\mathcal V\sinh y_2}&(\sinh y_2 \sinh \varphi - \cosh y_2 \cosh \varphi ) \\
 &=\frac{i\widetilde{k} x_\perp}{\mathcal V}\sqrt{n^2\mathcal V^2-1}\cosh(\varphi-y_2)\,.
\end{split}
\end{equation}
Considering the integral separately for positive and negative wavenumbers, we can write
\begin{align}
 I_2(\widetilde{k}>0) &= \frac{1}{i\pi}\int_{-\infty}^{\infty} d\phi \, e^{i \sigma_2 \cosh(\varphi-y_2)}, \\
 I_2(\widetilde{k}<0) &= \frac{1}{i\pi}\int_{-\infty}^{\infty} d\phi \, e^{-i \sigma_2 \cosh(\varphi-y_2)},
\end{align}
where the real and positive coefficient $\sigma_2$ is defined as
\begin{equation}
 \sigma_2 \equiv \frac{|\widetilde{k}|x_\perp}{\mathcal V}\sqrt{n^2\mathcal V^2-1} \, .
\end{equation}
The following integral representations of the Hankel functions~\cite{Abramhowitz:po1984} can now be applied
\begin{align}
 H_0^{(1)}(x) &= \frac{1}{i\pi}\int_{-\infty}^{\infty}dt \, e^{ix\cosh t}\,, \\
 H_0^{(2)}(x) &= -\frac{1}{i\pi}\int_{-\infty}^{\infty}dt \, e^{-ix\cosh t} \, ,
\end{align}
which hold for $x$ real and positive. Finally, the integral \req{eq:Iintegral} reads
\begin{equation}\label{eq:I2}
 I_2(\widetilde{k},x_\perp) = \Theta(\widetilde{k})H_0^{(1)}\left(\sigma_2\right) \\
 - \Theta(-\widetilde{k})H_0^{(2)}\left(\sigma_2\right)\,.
\end{equation}
for superluminal motion $n\mathcal{V} > 1$. 

\subsection{Overall solution for arbitrary $n\mathcal{V}$}
We can express the two regimes $I_1$ [\req{eq:I1}] and $I_2$ [\req{eq:I2}] of the integral $I(\widetilde{k},x_\perp)$ defined in \req{eq:Iintegral} as a single combination of Hankel functions
\begin{equation} \label{eq:solution}
\begin{split}
 I(\widetilde{k},x_\perp) &= \Theta(1-n\mathcal V)H_0^{(1)}(\xi) + \Theta(n\mathcal V-1)\\
 &\times \left[\Theta(\widetilde{k}) H_0^{(1)}(\xi) - \Theta(-\widetilde{k})H_0^{(2)}(\xi)\right] \,,
\end{split}
\end{equation}
where the argument of the Hankel functions $\xi$ was defined as
\begin{equation}\label{eq:xi}
 \xi \equiv \sigma_2 = i \sigma_1 = \frac{|\widetilde{k}|x_\perp}{\mathcal V}\sqrt{n^2\mathcal V^2-1} \,.
\end{equation}
This argument is real for superluminal velocities $n\mathcal V>1$ and imaginary for subluminal velocities $n\mathcal V<1$. 

\section{Electromagnetic field tensor}\label{app:EMtensor}

From the solution for the potential~\req{eq:app_potential}, the field tensor is 
\begin{equation}
 F^{\mu\nu}(x) = \partial^\mu A^\nu(x) - \partial^\nu A^\mu(x) \, .
\end{equation}
A word of caution: We will be exchanging a derivative with an integration, which is  mathematically allowed only when the field is sufficiently smooth. Singularities in fields that block the usual exchanging of operators can be closely related to Dirac strings~\cite{Jackson:1998nia}. In the context of Cherenkov radiation, it is known that singularities can appear for a constant index of refraction~\cite{Jackson:1998nia}, and when the particle crosses the Cherenkov threshold $n\beta=1$~\cite{Afanasiev:tr1998}. Remembering these reservations, we proceed: The partial derivative of the four-potential $A^\mu$ is equal to
\begin{equation}
\begin{split}
 \partial^\mu A^\nu(x) &= \frac{i\mu_0 q}{8\pi\mathcal{V}\Gamma} \int_{-\infty}^\infty d\widetilde{k}P_n^{\nu\alpha}u_\alpha e^{-i\widetilde{k}(\widetilde{x} - x_\parallel/\mathcal{V})}\\
 &\times \left[-i\widetilde{k} I(\eta^\mu/c - e_\parallel^\mu/\mathcal V) + \partial^\mu I\right] \,,
\end{split}
\end{equation}
where the four-gradient of $I$ is given by the four-gradient of $\xi$ defined in \req{eq:xi}
\begin{equation}\label{eq:partialmuI}
 \partial^\mu I(\widetilde{k},x_\perp) = \frac{\partial I}{\partial \xi}\partial^\mu \xi = \frac{\partial I}{\partial \xi}|\widetilde{k}|\frac{\sqrt{n^2\mathcal V^2-1}}{\mathcal V}e_\perp^\mu \,,
\end{equation}
and where the spacelike unit vectors $e_\parallel^\mu$, $e_\perp^\mu$ are defined in \reqs{eq:e_parallel}{eq:e_perp}. Introducing the four-vector $v^\mu$
\begin{equation}
 v^\mu(\widetilde{k}) \equiv P_n^{\mu\nu}(\widetilde{k})u_\nu \,,
\end{equation}
(note the definition of $P^{\mu\nu}_n$ in \req{eq:Pmn}), then the field tensor is
\begin{equation}\label{eq:app_emtensor}
\begin{split}
 &F^{\mu\nu}(x) = \frac{i\mu_0q}{4\pi\mathcal V\Gamma}\int_{-\infty}^\infty d\widetilde{k} \, e^{-i\widetilde{k}(\widetilde{x}-x_\parallel/\mathcal{V})}\\
 &\times \left[-i\widetilde{k} I \left(1-\frac{1}{n^2\mathcal V^2}\right)\frac{\eta^{[\mu} u^{\nu]}}{c}
 + \partial^{[\mu} (I) v^{\nu]}\right] 
\end{split}
\end{equation}
as a function of coordinates $\widetilde{x}$, $x_\parallel$, and the dependence on $x_\perp$ is contained in $I(\widetilde{k},x_\perp)$. 
\section{Displacement field tensor}\label{app:displacementtensor}
The displacement field tensor is defined as the convolution integral over the field tensor and the constitutive tensor density \req{eq:covariant_constitutive_eqn}
\begin{equation}\label{eq:app_constit}
 \mathcal H_{\mu\nu}(x) = \frac{1}{2}\int d^4x' \, \hat{z}_{\mu\nu\alpha\beta}(x-x')F^{\alpha\beta}(x') \,.
\end{equation}
The constitutive tensor density is for an isotropic medium described by an invariant dielectric constant density $\hat{\varepsilon}$ and an inverse vacuum permeability density $\hat{\mu_0}^{-1}$. Using the Bel expansion~\cite{bel2000radiation} we obtain \req{eq:constitutive_tensor} in the momentum space
\begin{equation}\label{eq:app_const_tensor}
\begin{split}
 \hat{z}^{\mu\nu\alpha\beta}(\widetilde{k}) &= -\hat{\varepsilon}(\widetilde{k}) (P^{\mu\beta}\eta^\nu \eta^\alpha + P^{\nu\alpha} \eta^\mu \eta^\beta \\ 
 &- P^{\mu\alpha}\eta^\nu\eta^\beta - P^{\nu\beta}\eta^\mu\eta^\alpha) \\ 
 &- \mu_0^{-1} g_{\delta\sigma}\epsilon^{\mu\nu\rho\sigma}\epsilon^{\alpha\beta\gamma\delta}\eta_\rho \eta_\gamma /c^2\, ,
\end{split}
\end{equation}
The inverse Fourier transformation of the constitutive tensor density can be decomposed into integrals over parallel and perpendicular components of $k^\mu$ with respect to $\eta^\mu$, see \req{eq:etadecompose}
\begin{equation}\label{eq:zint}
\begin{split}
 \hat{z}^{\mu\nu\alpha\beta}(x-x')= &\int \frac{d\widetilde{k}d^3\widetilde{k}_\perp}{(2\pi)^4}\hat{z}^{\mu\nu\alpha\beta}(\widetilde{k})\\
 &\times e^{-i\widetilde{k} (\widetilde{x}-\widetilde{x}')} e^{-i \widetilde{k}_\perp \cdot (x - x')}\,.
\end{split}
\end{equation}
The transverse momentum integral yields a 3D delta function
\begin{equation}\label{eq:zdelta}
\begin{split} 
 \hat{z}^{\mu\nu\alpha\beta}(x-x')&= \delta^3(\widetilde{x}_{\perp} - \widetilde{x}_{\perp}')\\
 &\times \int \frac{d\widetilde{k}}{2\pi}\hat{z}^{\mu\nu\alpha\beta}(\widetilde{k})e^{-i\widetilde{k} (\widetilde{x} - \widetilde{x}')}\, .
\end{split}
\end{equation}
Now, the same decomposition [\req{eq:etadecompose}] can also be applied to the position four-vectors to evaluate integrals in \req{eq:app_constit} $d^4x'=d\widetilde{x}'d^3\widetilde{x}'_\perp$. Inserting \req{eq:zdelta} into~\req{eq:app_constit}, we obtain
\begin{equation}
\begin{split}
 \mathcal H^{\mu\nu} &= \frac{1}{2}\int \frac{d\widetilde{k} }{2\pi}d\widetilde{x}'d^3\widetilde{x}'_\perp \, \hat{z}^{\mu\nu\alpha\beta}(\widetilde{k}) e^{-i\widetilde{k} (\widetilde{x}-\widetilde{x}')}\\
 &\times\delta^3(\widetilde{x}_{\perp} - \widetilde{x}_{\perp}') F_{\alpha\beta}(\widetilde{x}',\widetilde{x}_{\perp}')\\
 &=\frac{1}{2}\int \frac{d\widetilde{k} d\widetilde{x}'}{2\pi} \, \hat{z}^{\mu\nu\alpha\beta}(\widetilde{k}) e^{-i\widetilde{k} (\widetilde{x}-\widetilde{x}')}\\
 &\times F_{\alpha\beta}(\widetilde{x}',\widetilde{x}_{\perp})\,,
\end{split}
\end{equation}
where the spatial integration over transverse positions was performed. Substituting the position space electromagnetic tensor \req{eq:app_emtensor} yields
\begin{equation}
\begin{split}
 \mathcal H^{\mu\nu}&= \frac{i\mu_0 q}{16\pi^2 \Gamma \mathcal{V}}\int d\widetilde{k} d\widetilde{k}' d\widetilde{x}' \, \hat{z}^{\mu\nu\alpha\beta}(\widetilde{k})\\
 &\times e^{-i\widetilde{k} (\widetilde{x}-\widetilde{x}')}e^{-i\widetilde{k}' (\widetilde{x}'-x'_\parallel/\mathcal{V})}\\
 &\times\left[-i\widetilde{k}' I' \left(1-\frac{1}{n'^2\mathcal V^2}\right)\frac{\eta_{[\alpha} u_{\beta]}}{c} + \partial_{[\alpha} (I') v'_{\beta]}\right] \, .
\end{split}
\end{equation}
where $I'$, $n'$, $v'_\mu$ are functions of $\widetilde{k}'$. Additionally, $I'$ is a function of $x_\perp'$ but note that both coordinates $x_\parallel$ and $x_\perp$ from \reqs{eq:x_parallel}{eq:x_perp} are only functions of the transverse position $\widetilde{x}_{\perp}^\mu$ which we have already integrated over. Also,
\begin{equation}
 \widetilde{x}' - \frac{x'_\parallel}{\mathcal{V}} = \widetilde{x}' + \frac{\widetilde{x}_{\perp}\cdot u}{\mathcal{V}^2\Gamma c} = \widetilde{x}'-\frac{x_\parallel}{\mathcal{V}} 
\end{equation}
and the integral over $d\widetilde{x}'$ now separates and yields $2\pi \delta(\widetilde{k} - \widetilde{k}')$. Applying this delta function to integrate over $d\widetilde{k}'$ 
\begin{equation} \label{eq:displacement_tensor2}
\begin{split}
 &\mathcal H^{\mu\nu} = \frac{i\mu_0 q}{8\pi\mathcal{V}\Gamma}\int_{-\infty}^\infty d\widetilde{k} \, \hat{z}^{\mu\nu\alpha\beta}(\widetilde{k}) e^{-i\widetilde{k}(\widetilde{x}-x_\parallel/\mathcal{V})}\\
 &\times\left[-i\widetilde{k} \left(1-\frac{1}{n^2\mathcal V^2}\right)\frac{\eta_{[\alpha} u_{\beta]}}{c} + \partial_{[\alpha} (I) v_{\beta]}\right] \,.
\end{split}
\end{equation}
Finally, inserting the expression for the constitutive tensor density~\req{eq:app_const_tensor} into~\req{eq:displacement_tensor2}, we obtain
\begin{equation} \label{eq:displacement_tensor3}
\begin{split}
 &\mathcal H^{\mu\nu} = \frac{i\mu_0q}{8\pi\mathcal{V}\Gamma}\int_{-\infty}^\infty d\widetilde{k} \, e^{-i\widetilde{k}(\widetilde{x}-x_\parallel/\mathcal{V})}\left[-\hat{\epsilon}(P^{\mu\beta}\eta^\nu\eta^\alpha \right.\\
 &+ P^{\nu\alpha}\eta^\mu\eta^\beta - P^{\mu\alpha}\eta^\nu\eta^\beta - P^{\nu\beta}\eta^\mu\eta^\alpha)\\
 &\left. -\mu_0^{-1} g_{\delta\sigma}\epsilon^{\mu\nu\rho\sigma}\epsilon^{\alpha\beta\gamma\delta}\eta_\rho\eta_\gamma/c^2\right]\\
 &\times \left[-i\widetilde{k} I\left(1-\frac{1}{n^2\mathcal V^2}\right) \frac{\eta_{[\alpha} u_{\beta]}}{c} + \partial_{[\alpha} (I) v_{\beta]}\right] \,.
\end{split} 
\end{equation}
Tensor contractions in~\req{eq:displacement_tensor3} need to be computed to simplify. First, we evaluate
\begin{align}
 &(P^{\mu\beta}\eta^\nu\eta^\alpha + P^{\nu\alpha}\eta^\mu\eta^\beta - P^{\mu\alpha}\eta^\nu\eta^\beta - P^{\nu\beta}\eta^\mu\eta^\alpha) \\ \nonumber
 &\times \eta_{[\alpha} u_{\beta]} = -2c^2\eta^{[\mu} u^{\nu]} \,,
\end{align}
and second
\begin{align}
 \notag&(P^{\mu\beta}\eta^\nu\eta^\alpha + P^{\nu\alpha}\eta^\mu\eta^\beta - P^{\mu\alpha}\eta^\nu\eta^\beta - P^{\nu\beta}\eta^\mu\eta^\alpha) \\
 &\times\partial_{[\alpha} (I) v_{\beta]} = -\frac{2\Gamma c^2}{n^2}\partial^{[\mu}(I) \eta^{\nu]}\, .
\end{align}
The last non-vanishing product is 
\begin{equation}
\begin{split} g_{\delta\sigma}&\epsilon^{\mu\nu\rho\sigma}\epsilon^{\alpha\beta\gamma\delta}\eta_\rho\eta_\gamma \partial_{[\alpha} (I) v_{\beta]} / c^2 \\
&= -2\partial^{[\mu}(I)u^{\nu]} + 2 \Gamma \partial^{[\mu}(I)\eta^{\nu]}\,.
\end{split}
\end{equation}
Finally, considering $\mu_0 \hat{\varepsilon}(\widetilde{k}) = n^2(\widetilde{k})/c^2$, the result of the displacement field tensor calculation is
\begin{equation}
\begin{split}
 &\mathcal H^{\mu\nu}(x) = \frac{iq}{4\pi\mathcal V\Gamma}\int d\widetilde{k} \, e^{-i\widetilde{k}(\widetilde{x}-x_\parallel/\mathcal{V})}\\
 &\times\left[-i\widetilde{k} I n^2\left(1-\frac{1}{n^2\mathcal V^2}\right)\frac{\eta^{[\mu} u^{\nu]}}{c} + \partial^{[\mu}(I) u^{\nu]}\right]
\end{split}
\end{equation}
as a function of coordinates $\widetilde{x}$, $x_\parallel$, and the dependence on $x_\perp$ is contained in $I(\widetilde{k},x_\perp)$. 

To verify this result, from the corresponding components of $F^{\mu\nu}$ and $\mathcal H^{\mu\nu}$ it can be shown that in the medium rest frame and for constant $n^2(\widetilde k)$ the following relations hold
\begin{align}
 \boldsymbol D(x) &= \frac{n^2}{\mu_0 c^2} \boldsymbol E(x) = \varepsilon \boldsymbol E(x)\\ 
 \boldsymbol H(x) &= \frac{1}{\mu_0}\boldsymbol B(x) \,,
\end{align}
which is consistent with the usual three-dimensional constitutive relations.
\section{Field momentum change due to hypercylinder caps}\label{app:cylinder_caps}
A similar integration of the field momentum change as was performed in Section~\ref{sec:RR} for the curved hypercylinder surface can also be computed for the hypercylinder caps. The surface elements for the two caps are
\begin{equation}
 d^3\sigma_{\nu\pm} = \pm g_{\nu\alpha}e^\alpha_\parallel x_\perp dx_\perp d\phi d\widetilde{x}\,,
\end{equation}
while the coordinate $x_\parallel$ is set to a constant $\mp L$. Then the change of field momentum for each cap is
\begin{equation}
 dp^\mu_{\text{EM}\pm} = \pm \frac{1}{c} \int x_\perp d\phi dx_\perp d\widetilde{x}\, T^{\mu\nu}e_{\parallel\nu} \,.
\end{equation}
Using the same parameterization for the energy momentum tensor as in \req{eq:kernel} 
\begin{equation}
\begin{split}
 dp^\mu_{\text{EM}\pm} &= \pm \frac{1}{c}\int x_\perp d\phi dx_\perp d\widetilde{k} d\widetilde{k}' d\widetilde{x}\\
 &\times e^{-i(\widetilde{k}-\widetilde{k}')(\widetilde{x}\pm L/\mathcal V)} \mathcal{T}^{\mu\nu}(\widetilde{k},\widetilde{k}',x_\perp)e_{\parallel\nu}\, .
\end{split}
\end{equation}
At this point, the integration over the coordinate $d\widetilde{x}$ can be performed, yielding $2\pi\delta(\widetilde{k}-\widetilde{k}')$, making the integration over $d\widetilde{k}'$ feasible. Therefore,
\begin{equation}
 dp^\mu_{\text{EM}\pm} = \pm \frac{2\pi}{c} \int x_\perp d\phi dx_\perp d\widetilde{k}\, \mathcal{T}^{\mu\nu}(\widetilde{k},\widetilde{k},x_\perp)e_{\parallel\nu}
\end{equation}
and the dependence on the cap positions vanished. The contribution from both caps is then equal and opposite and adds up to zero without the need to explicitly evaluate. 
\normalem 
%

\end{document}